# ON THE VOLTAGE-CONTROLLED ASSEMBLY OF NP ARRAYS AT ELECTROCHEMICAL SOLID/LIQUID INTERFACES

Cristian Zagar[1], Ryan-Rys Griffith[1], Rudolf Podgornik[2], and Alexei A. Kornyshev[1,3]*

[1] Department of Chemistry, Imperial College London, SW7 2AZ London, United Kingdom

[2] School of Physical Sciences, University of Chinese Academy of Sciences, and Kavli Institute of Theoretical Sciences, No.380 Huaibei, Huairou District, Beijing 101408, P. R. China

[3] Thomas Young Center for Theory and Simulation of Materials, SW7 2AZ London, United Kingdom

*a.kornyshev@imperial.ac.uk

Research in the field of nanoplasmonic metamaterials is moving towards more and more interesting and, potentially useful, applications. The present work tackles the problem of nanoparticle self-assembly at an electrochemical solid-liquid interface from a purely theoretical perspective. We perform a simplified, comprehensive analysis of the stability of a nanoparticle arrays under different conditions and assembly. From the Poisson-Bjerrum model of electrostatic interactions between a metallic nanoparticle and the electrode and between the nanoparticles at the electrode, as well the Hamaker-Lifshitz model of the corresponding van der Waals interactions, we reach some conclusions regarding the possibility to build arrays of charged nanoparticles on electrodes and disassemble them, subject to variation of applied voltage. Since system of this type have been shown, recently, to provide nontrivial electrotuneable optical response, such analysis is crucial for answering the question whether such scenarios of electrochemical plasmonics are feasible.

The term *optical metamaterials* is attributed to materials with odd, disruptive, often counterintuitive optical properties, the principles of operation of which are often based on subtle, nontrivial physical electromagnetic effects in nanoscale structures. An explosive development of this area was made possible with the progress in nanotechnology. Indeed, the ability to design nanostructures which control electromagnetic wave propagation revealed the potential for many interesting and useful applications, from sensors and optical switches[1] to superlenses and communication technology[2,3]. Many of such designed structures are built, however, fixed, and the demand to build *tuneable* metamaterials developed mainly in the investigation of electromagnetic effects that may cross-influence each other without changing the material's nano-structure[4–10]. If successful, switching the properties could be very fast. A much simpler approach would be to tune optical signals through changing the structure of the material. Well known are the attempts to do this mechanically; c.f. the material named plasmene, in which an array of metallic 'plasmonic' **n**ano**p**articles (NPs) is chemisorbed on a ribbon and stretching the ribbon one could change the optical response of the array. Generally, assembling NPs into arrays or, even more complex structures[11], leads to interesting reflectance and transmission spectra, resulting from plasmon resonances induced by light, and a way towards tuning the optical response of such systems is to tune their structures.

One way of tuning the structure of plasmonic NP arrays in real time is their voltage controlled self-assembly/disassembly at electrochemical interfaces. The idea first was proposed in Ref[12]. and later developed in a series of papers (for review see ref.[13]). Its first experimental realization has been presented in Ref.[14] for an electrochemical liquid-liquid interface, backed up by the theory of optical response spectra from such systems[15]. A detailed theory of optical spectra from the arrays of NPs at solid electrode-liquid electrolyte interfaces was presented in Ref.[16], where the presence of plasmonic substrate changes dramatically the character of the signal. Its first experimental realisation has been reported just now[17], in full harmony with theory. Let us summarize briefly the main idea of these works.

In an electrochemical liquid-liquid cell, at the interface of two immiscible electrolytic solutions[18] (say NaCl in water and TBA-TPB in 1,2, dichloroethane), plasmonic NPs, say AuNp-s adsorb at the interface, piercing it, to block the unfavourable surface between water and oil. If the surface energy of a NP in contact with water and with oil is lower than the surface energy of that blocked interface area, they will form a so-called capillary well that will keep NP at the interface[19], and NPs which are dissolved exclusively in water get adsorbed. But NPs from the very beginning are functionalized by ligands the head groups of which dissociate in water and leave ligands charged – for mercaptanoid acid, negatively charged. Functionalization is needed to ensure colloidal stability of NPs in aqueous electrolyte bulk, otherwise they will fuse due to Van der Waals forces. The strength of thus provided electrostatic repulsion is controlled by two factors: (i) electrolyte concentration (inorganic electrolyte in water and organic electrolyte in oil) – the higher concentration provides the Debye screening of electrostatic interactions; (ii) the net charge of ligands on the NP (usually controlled through pH). But somehow, since we made NPs repelling each other, they will tend not to sit close to each other when adsorbed at the interface. And this is exactly what we see experimentally, either through X-ray characterisation of the NP arrays at the interface, or through their optical reflectivity[20] (the reflection is stronger for denser arrays, and the overall spectrum changes: maximum reflectivity shifts to the red in full accordance with the developed theory[15]). In order to make the array denser for a given pH and electrolyte concentration, we need to increase the driving force for each particle to get adsorbed at the interface. One way to do it, is to apply voltage across the interface in an electrochemical cell, i.e. polarise aqueous phase more negatively than oil. This was shown to be perfectly reversible, allowing voltage-controlled formation of NP arrays at liquid-liquid interfaces, and thereby building the first electro-tuneable/switchable mirror[14].

Liquid/liquid electrochemical interfaces have a lot of interesting features and advantages, but since it is hard to maintain those interfaces vertical, solid-liquid systems are of special interest. These can be of two kinds:

1. Solid transparent electrode [e.g., Indium Tin Oxide (ITO)] in contact with aqueous electrolyte solution.

2. Metal electrode in contact with aqueous electrolyte solution.

The first class of systems will function similar to the liquid|liquid one. Polarizing electrode positively, it will favour negatively charged NPs to get adsorbed at the interface to form a dense array and thereby provide a mirror function; polarizing the electrode negatively will push NPs away, into the bulk, and will make the interface transparent.

The second class of systems behave entirely differently, and in two possible ways, depending on the material of the substrate metal and of the NP. First of all, when NPs are not adsorbed on the solid substrate, the latter is a mirror. If it is gold, the mirror is not perfect, but having adsorbed a homogeneous array of AgNp-s make its reflection spectrum flatter, more perfect. If the substrate is silver, and NPs are AuNp-s, the effect is opposite, always a perfect mirror gets a broad dip in reflection spectrum, i.e. mirror is acquiring a colour, and the denser the array of AuNPs, the stronger the dip and the corresponding colour change, as predicted by the theory[16] and confirmed by experiment, in full agreement with the theory[17].

More systems of this kind can be envisaged[13,21], but all can be categorized as *electrochemical plasmonics* (EP) systems. Few details on solid/liquid EP-systems should be mentioned before we focus on the subject of this paper.

The speed of response to changed voltage, so far demonstrated was very slow, but… expectedly slow. Indeed, the capillary well at the interface extends just over the Debye length in electrolyte. So, if particles are left dispersed in the bulk of a macroscopic solution and in amounts to just cover the interface, it may take hours for them to randomly diffuse toward the surface and get trapped in the well. When, however they leave the well, the array loses its coherence very fast, and the mirror function disappears. The kinetics of NP adsorption in macroscopic systems have been experimentally studied for both liquid|liquid[14] and solid|liquid[17] systems, in full agreement with theoretical expectations. It was made clear that if the adsorption kinetics is fully controlled by diffusion of NPs from the bulk, there is a very simple recipe how to speed it up: the time for reaching the interface is roughly inversely proportional to the square of NP concentration. The way to increase the latter without making the solution coloured was to increase the thickness of the aqueous phase. That effect has been studied and the one over square root of concentration law was approved. Thus, the way to reach a millisecond response time was straightforward: build a micro-or nanocell.

It remained, however, to be understood, under which conditions adsorption and desorption of NPs are diffusion controlled, and what is basically the balance of forces that brings NPs to or pushes them away from a neutral or polarized electrode. Next, we need to understand how such NPs would interact with each other at the interface, and what the coverage dependence on applied voltage, concentration of electrolyte and the charge of functionalized NPs will be in the end. Whereas for liquid|liquid systems a more primitive theory of this kind has been developed,[22,12] which justified the principle possibility of electrovariable plasmonics, this has not been done for solid|liquid interfaces. The latter task is the main subject of the present paper. It attempts to give a first theoretical basis on whether voltage-control over the density of the adsorbed NP arrays at solid electrodes is, in principle, possible, which is the foundation of electrochemical plasmonics at solid electrodes. Such a study is also expected to reveal the means for the fine tuning of such control, through adjusting electrolyte concentration and ionization of ligands, as well as explore the effect of the size of NPs.

Note that generally the theoretical machinery of electrochemical plasmonics is comprised of three main components.

1. The theory of stability of NPs arrays at a polarised electrode, characterised by an equilibrium electrosorption isotherm, based on the theory of interactions of NPs with the electrode and with each other.

2. The theory of NP adsorption/desorption kinetics, based on a quasi-steady state approximation for diffusion and an adsorption isotherm for surface coverage that has common elements with the theory of adsorptions kinetics of (macro)molecules.

3. Electrodynamic multilayer stack model, which can quickly provide the optical response of NP arrays assembled near (generally, film-covered) substrates, for a given structure of the array, size, shape, and material of NPs, and their disposition with respect to the substrate.

The third component is well developed[15,16], giving excellent results as compared to numerical COMSOL simulations, but taking seconds to get the full spectra, with the transparency of results, which allows to avoid the 'black-box' simulations. The second component can be based on the adjustment[17] of existing kinetic theory of adsorption[23]. But the first component is the least, if at all developed, and the present study makes the first steps in this direction.

The last comment before we begin is that we will consider different electrodes, transparent ITO type, or metallic, like gold and silver, but considering the latter we may need to assume a protective layer on them, such as e.g. TiN, of SAMs, which are often use to avoid oxidation of surfaces, or passivate the electrode against water electrolysis or electrochemical reactions of ions of electrolyte for the applied electrode potentials.

# 1. Electrostatic vs Van der Waals forces

Two main effects will be accounted for: electrostatic and Van der Waals interactions.

As mentioned previously, metallic NPs in solution always exhibit attractive van der Waals interactions, which are strong enough to force them to agglomerate into clusters. Stabilising the solution is, therefore, crucial, and is achieved by functionalising NPs with ligands that can ionise. Usually, these ligands are mercaptanoic acids, which lose protons from their carboxyl groups and become negatively charged. Ionisation of ligands, therefore, attempts to create enough charge on the surface of the particles that electrostatic repulsions stop them from aggregating. Under these conditions, the charge on NPs can be adjusted by changing only two 'chemical' parameters. First, pH is what controls the fraction of dissociated ligand molecules directly. By increasing the pH, the number of ionised ligand molecules also increases, so NPs have more charge around them and repulsion becomes stronger. The second parameter is electrolyte concentration. The higher the concentration, the weaker the electrostatic repulsion becomes. Usually the balance between pH, electrolyte concentration and, sometimes NP concentration as well, is found experimentally, and there is not much flexibility left in these parameters once the solution is prepared.

Description of electrostatic interaction of NPs in solution near the interface with the interface and each other is a tricky task, as it involves the response of the metal substrate, i.e. image forces also screened by electrolyte ions. Furthermore, when the electrode is polarized, an electrical double layer will be formed at the interface, and the electric field of the double layer will act on the charges of NPs. We will explore the simplest possible approximation to the solution of this problem, considering those charges fixed, as well as ignore the polarizabilty of the particles. Note, furthermore, that will not be involved here in more complicated theory that allow for like-charge repulsion, because we will not be considering electrolytic solutions with large Bjerrum lengths, dealing exclusively with 1-1 aqueous electrolytes, as experimentally most practical in electrochemical plasmonics[14,17]; thus, electrostatic interactions between nanonoparticles that are charged in the same way will be solely repulsive.

When considering Van der Waals interactions of the particles with electrodes we will use standard expressions of the Lifshitz theory[24]. Considering interaction between NPs we will make an estimate of the largest possible effect, ignoring the effect of the metallic substrate. The theory of interaction of two metallic spheres of finite radius near a flat metallic substrate is cumbersome and not fully developed, but from the theory of point-like fluctuating dipoles near metal substrate[25,26,27] we know the effect of such substrate will be in reducing the Van der Waals attraction.

All these calculations will be performed to figure out (within the mentioned theoretical framework) whether spontaneous assembly or disassembly can be induced by changing voltage and, furthermore, to show how the surface NP population responds to its change.

# 2. Model assumptions and basic equations

The simplest and most natural way of describing a solid-liquid interface is by modelling it as a plane which separates two semi-infinite media, namely an aqueous electrolyte and a solid metallic or semi-metallic material. Geometrically, this approach works under two conditions, that should be fulfilled in practice. First, the quasi-flat approximation is adequate only if the surface roughness is small. In this case, small implies it is practically flat down to the nanoscale. Second, the semi-infinite description is reliable only if the electric fields present in the system do not reach the physical end of the solid or liquid phases. Theoretically, this happens if the characteristic screening lengths in the two phases are short compared to the size of the system. For the aqueous phase, electrolyte concentrations are typically within 10-100 mM, leading to Debye screening lengths of the order of nm, which stop electric fields from propagating towards the physical boundaries of the system. Regarding the solid phase, electrons tend to screen the electric fields very efficiently. A simple estimate of this capability can be done with the help of the so-called Thomas-Fermi screening theory. For a metal, its value is extremely small, of the orders of A, or even less. This is caused by the very loose binding of the electrons in the conduction band and it makes those electrons move almost freely within the structure. For a semi-metal, electrons are bound more tightly, leading to an increase in Thomas-Fermi length to the order of a few nm. Even in this case, unless we deal with an electrochemical 'nano-cell'[28], the system remains big enough to screen the fields completely.

Below, the main types of interactions – electrostatic and van der Waals, are treated independently, and their contributions added towards the overall effect. Considering this is a verified approach for soft interfaces[29], it also should, in principle, give at least an estimate of the energies present in the system. In the following sections, two phases, aqueous and solid, will be named phase 1 and phase 2, respectively, and the variables and constants associated with them will be labelled accordingly. Full derivations of the electrostatic interactions are also given in Appendix 1.

## 2.1 Electrostatic interactions

The most common way of modelling electric potentials in electrolytes and electrolyte-like systems is the Poisson-Boltzmann equation[30]:

$$\nabla^2 \phi + \sum_i \frac{z_i \nu_i e}{\epsilon_0 \epsilon_1} exp\left(-\frac{z_i e \phi}{kT}\right) = -\frac{\rho}{\epsilon_0 \epsilon}, \qquad (1)$$

where $\phi$ is the electric potential, as a function of coordinates, $z_i$-the valence of ion $i$, $\nu_i$-number of ions per molecule of electrolyte (i.e., stoichiometric coeffecient), $e$ is the elementary charge, $kT$ is thermal energy, $\epsilon_0$ and $\epsilon$ are the permittivity of the vacuum and dielectric constant respectively, and $\rho$ is 'free charge' that we will associate with NPs.

The nonlinearity of this equation generates great difficulty in solving it for the complex geometry consisting of spherical NPs interacting with a charged interface. Although numerical solutions can be obtained, the possibility of extracting analytical expressions is still preferable because of the intuition one can develop about the dominating effects. One way around the problem of nonlinearity is to use linear approximation on the exponentials in eq. (1) resulting in the linear Poisson-Boltzmann equation.

$$\nabla^2 \phi - k^2 \phi = -\frac{\rho}{\epsilon_0 \epsilon}, \quad (2)$$

where $k$ represents the inverse Debye screening length, given by

$$k = \sqrt{\frac{\nu_i z_i^2 e^2}{\epsilon_0 \epsilon k T}} \quad (3)$$

Even within the linear regime, calculations proved to be very cumbersome, but full derivations are given in the appendix.

The difference in electrostatic properties between the two phases results in two mathematical solutions, one on each side of the interface. They have to match the boundary conditions at the interface, the continuity of electric potential and continuity of the normal component of electric induction. Apart from dielectric properties, another important difference is that the electrolyte phase contains free charge, in this case in the form of ionised ligands on the surface of NPs. If $\epsilon_1$ and $\epsilon_2$ are dielectric constants of the electrolyte and solid phase, and $k_1^{-1}$ and $k_2^{-1}$ their respective screening lengths, the potential obeys the following equations:

$$\nabla^2 \phi_1 - k_1^2 \phi_1 = -\frac{\rho}{\epsilon_0 \epsilon_1} \quad (4)$$

$$\nabla^2 \phi_2 - k_2^2 \phi_2 = 0 \quad (5)$$

Here the two functions, $\phi_1$ and $\phi_2$ are expressions for the electric potential in electrolyte and solid electrode, respectively. As the solid phase does not contain any free charge, the free term in the second equation is zero. The situation when the interface has free charge (because of a change in electrode potential from the potential of zero charge) will be treated separately.

Mathematically, the two boundary conditions can be expressed in cylindrical coordinates as:

$$\phi_1(\vec{R}, z = 0) = \phi_2(\vec{R}, z = 0) \quad (6)$$

$$\epsilon_1 \frac{\partial \phi_1}{\partial z}\bigg|_{z=0} = \epsilon_2 \frac{\partial \phi_2}{\partial z}\bigg|_{z=0} \quad (7)$$

The solutions for $\phi_1$ and $\phi_2$ can be calculated by using the Fourier transform method. In short, the functions can be written in terms of the Fourier transforms, leading to the reciprocal versions of both equations (4) and (5) and boundary conditions.

The calculations lead to an expression for Fourier transforms of the two potentials in terms of the Fourier transform of the charge density. The result is, therefore, general enough to be used for any free charge distribution occurring in the electrolyte.

$$\tilde{\phi}_1\left(\vec{K},z\right) = \frac{1}{2\epsilon_0\epsilon_1\sqrt{K^2+k_1^2}}\left(\int_0^\infty e^{-\sqrt{K^2+k_1^2}|z-z_0|}\tilde{\rho}\left(\vec{K},z_0\right)dz_0 + \frac{\epsilon_1\sqrt{K^2+k_1^2}-\epsilon_2\sqrt{K^2+k_2^2}}{\epsilon_1\sqrt{K^2+k_1^2}+\epsilon_2\sqrt{K^2+k_2^2}}\int_0^\infty e^{-\sqrt{K^2+k_1^2}(z+z_0)}\tilde{\rho}\left(\vec{K},z_0\right)dz_0\right) \quad (8)$$

$$\tilde{\phi}_2\left(\vec{K},z\right) = \frac{\int_0^\infty e^{\sqrt{K^2+k_2^2}z - \sqrt{K^2+k_1^2}z_0}\tilde{\rho}\left(\vec{K},z_0\right)dz_0}{\epsilon_0\left(\epsilon_1\sqrt{K^2+k_1^2}+\epsilon_2\sqrt{K^2+k_2^2}\right)} \quad (9)$$

From $\tilde{\phi}_1$ and $\tilde{\phi}_2$ the potential energy of a charge distribution in front of an interface can be written as an integral of these functions over the charge distribution.

$$W = 2\pi^2 \int_0^\infty dz \int d\vec{K}\, \tilde{\phi}\left(\vec{K},z\right)\tilde{\rho}\left(-\vec{K},z\right) \quad (10)$$

Apart from the possibility of getting analytical expressions, the linear Poisson-Boltzmann equation also offers a strategical advantage. If the free charge distribution is separated into multiple pieces, the charge density generating the electric field will be the sum of the charge densities of the pieces.

$$\rho = \sum_i \rho_i \quad (11)$$

One can see both from the linear Poisson-Boltzmann equation and from solutions (8) and (9) that the potential depends linearly on the charge density. In other words, the total potential can be written as a sum of the potentials that each free charge domain would generate, independently.

$$\phi = \sum_i \phi_i \quad (12)$$

This also allows the electric field and energy of the system to be separated into multiple contributions. Of course, the validity of this superposition principle is totally based on the linearization of the Poisson-Boltzmann equation: in non-linear theory one cannot decouple different Fourier -harmonics.

For a two-dimensional NP array in front of the interface, the total energy can be divided as follows: image potential energy of one particle, energy of one particle interacting with the free charge on the electrode, and pair interaction energy between particles. When it comes to interacting particles, the pair interaction is the key quantity, as interaction with multiple particles can be written as a pairwise summation of interaction energies.

It is also important to note that $\tilde{\phi}_1$, given in eq (8) contains two terms. The first term represents the potential of the charge distribution in the bulk electrolyte. The energy given by this term is, by definition, the energy required to build the charge distribution given by $\tilde{\rho}$ out of point charges brought from an infinitely large distance. More importantly, the contribution to the energy determined by this first term does not depend on the distance of the charge distribution from the interface. An easy way to visualize it is by mathematically eliminating the interface ($\epsilon_1 = \epsilon_2$ and $k_1 = k_2$). Such operation affects only the second term (eliminating it!), which means it is the second term that is responsible for the effect of the interface on the potential of the charge distribution and, implicitly, on its energy. This term will be labelled $\delta\tilde{\phi}_1$, and the associated image energy, $\delta W$.

$$\delta W = 2\pi^2 \int_0^\infty dz \int d\vec{K}\, \delta\tilde{\phi}\left(\vec{K},z\right) \tilde{\rho}\left(-\vec{K},z\right) \tag{13}$$

### 2.1.1 Interaction of one particle with the interface

For this specific interaction, free charges come only from the ionised ligands on the surface of one particle. The simplest model to mimic that charge distribution is to consider it homogeneously distributed over an infinitely narrow spherical shell, which implies no electric field inside the shell. In reality, the charge discreteness allows the electric field to penetrate beyond the ligand carboxyl groups and also into the metal, but the separation between the ligands is sufficiently small. They are also undulating, smearing the effect of discreteness. Of course, if only a small part of ligands is ionized, this approximation may not be accurate, but this case would not be too interesting, as NPs would be prone to aggregation in the bulk. Still, in the case of acidic ligands, on average, there will be no isolated ionised regions on the particle, as the rate of proton rearrangement across the entire surface is very fast.

To write the charge density mathematically, one needs to look at the geometry of the system first. Let us consider a spherical shell, of radius $a$, which has its centre located at a distance $z_0$ from the interface, as shown in fig 1. If the total charge is $q$, then the charge density in cylindrical coordinates is

$$\rho(R,z) = \frac{q}{4\pi a^2} \delta\left(\sqrt{R^2 + (z-z_0)^2} - a\right), \tag{14}$$

with a corresponding Fourier transform

$$\tilde{\rho}(K,z) = \frac{q}{8\pi^2 a} J_0\left(K\sqrt{a^2 - (z-z_0)^2}\right) \Theta(a - |z - z_0|) \tag{15}$$

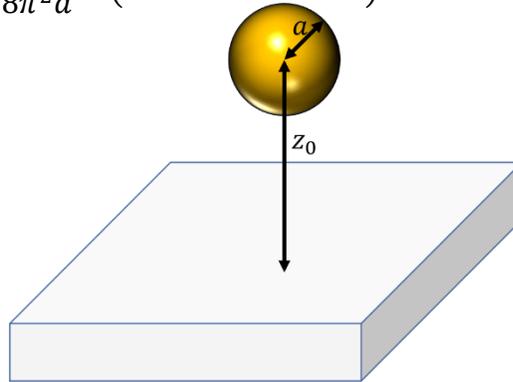

Fig.1 One NP of radius $a$, at a distance $z_0$ from the electrode

Substituting the charge density into the expressions for image potential $\delta \widetilde{\phi}_1$, and then into eq (13) gives the image potential energy as

$$\frac{\delta W}{kT} = N^2 \frac{L_B}{2} \left(\frac{sinh(k_1 a)}{k_1 a}\right)^2 \int_0^\infty \frac{KdK}{\sqrt{K^2+k_1^2}} \frac{\epsilon_1\sqrt{K^2+k_1^2} - \epsilon_2\sqrt{K^2+k_2^2}}{\epsilon_1\sqrt{K^2+k_1^2} + \epsilon_2\sqrt{K^2+k_2^2}} e^{-2\sqrt{K^2+k_1^2}z_0}, \quad (16)$$

where $N$ is the number of elementary charges on the particle, $L_B$ is Bjerrum length ($=e^2/4\pi\epsilon_0\epsilon_1 kT$) in the electrolyte, and $kT$ is thermal energy. The structure of this formula coincides with the general expression for image energies at metal-electrolyte interfaces. A very simple limiting case to test this formula is the point charge. Within this limit, the radius of the sphere considered above becomes infinitely small ($a \to 0$), leading to the already known image energy of a point charge near the interface of two plasma-like media[31].

$$\frac{\delta W}{kT} = N^2 \frac{L_B}{2} \int_0^\infty \frac{KdK}{\sqrt{K^2+k_1^2}} \frac{\epsilon_1\sqrt{K^2+k_1^2} - \epsilon_2\sqrt{K^2+k_2^2}}{\epsilon_1\sqrt{K^2+k_1^2} + \epsilon_2\sqrt{K^2+k_2^2}} e^{-2\sqrt{K^2+k_1^2}z_0} \quad (17)$$

The attractive or repulsive nature of the image force depends directly on the dielectric properties of the two phases. Usually it is a combination of both, resulting in a minimum located closer or farther away from the interface, depending on the dielectric constants and screening lengths of the two media. This behaviour comes from the change in sign caused by the middle factor, $\frac{\epsilon_1\sqrt{K^2+k_1^2} - \epsilon_2\sqrt{K^2+k_2^2}}{\epsilon_1\sqrt{K^2+k_1^2} + \epsilon_2\sqrt{K^2+k_2^2}}$. The specific situation where the image force is attractive at all distances is the ideal metal, with a diverging dielectric constant $\epsilon_2 \to \infty$. After integration, eq (17) gives

$$\frac{\delta W}{kT} = -N^2 L_B \frac{e^{-2k_1 z_0}}{4z_0} \quad (18)$$

An illustration of this special case is given in fig 2 a).

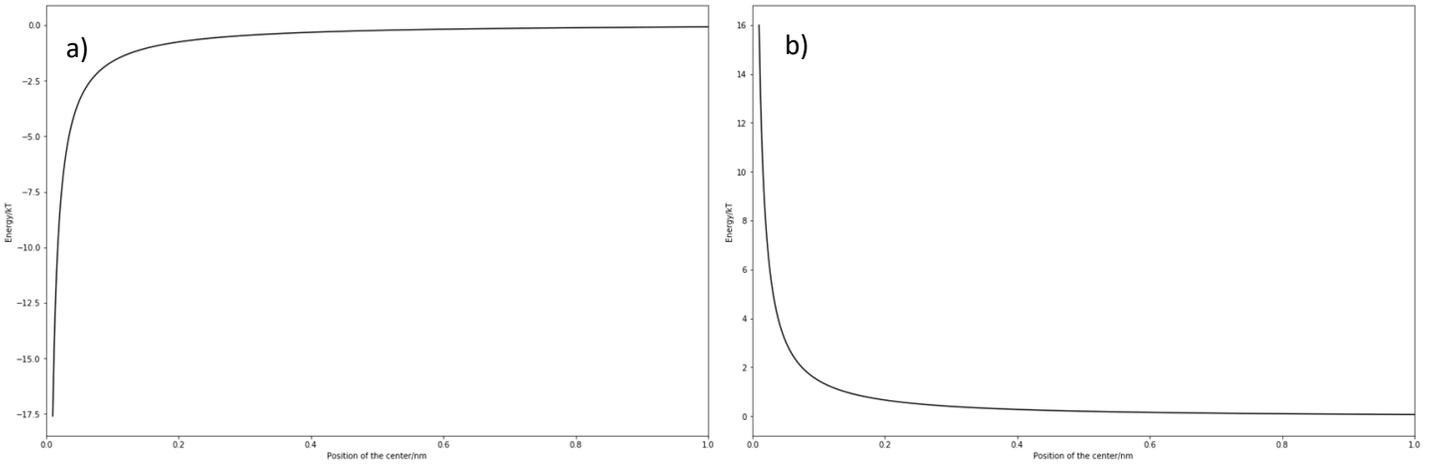

Fig. 2. Image potential energy of a point charge N=1 near a) an ideal metal plate (with $L_B = 0.7$ nm, $k_1 = 0.4$ nm$^{-1}$) and b) an ITO plate (with $\epsilon_1 = 79, k_1 = 0.4, \epsilon_2 = 3.62, k_2 = 1.67$ nm$^{-1}$)

When the electrode is made of a semi-metal, the image force becomes repulsive when the point charge is very close to the interface ($z_0 \to 0$), because such a material usually has a much lower dielectric constant than the electrolyte solution ($\epsilon_1 \gg \epsilon_2$). An example can be seen in fig. 2 b) for a dilute electrolyte (1 mM).

### 2.1.2 Interaction of one particle with the charge on the electrode

Depending on the voltage, the surface charge density on the electrode can be so high that significant nonlinear screening effects can come into play. Even if that happens to NPs as well, the problem can be circumvented in that case by renormalizing the charge on the particle. This approximation relies on the fact that most of the charge, which is screened nonlinearly within a very narrow range around the particle, does not contribute to the far field solution. In other words, only a small portion, an effective charge, contributes to both particle-electrode and pair interactions as long as the distances are not extremely short (3 nm). Such approximation is called sometimes, the 'Debye-Bjerrum' approximation. The calculation of this effective charge will be discussed later.

For the planar electrode, however, there is an analytical solution of the Poisson-Boltzmann equation for a 1:1 electrolyte.

$$V(z) = \frac{2kT}{e} \ln \frac{\coth \frac{eV_0}{4kT} + e^{-k_1 z}}{\coth \frac{eV_0}{4kT} - e^{-k_1 z}}, \tag{19}$$

where $V_0$ is the difference of the electrode potential from the potential of zero charge. In the approximate formula derived below, we want to take into account that the potential inside the metallic NP is constant, but to derive an analytical formula we simplify the derivation by assuming that the potential at any 'altitude' inside the NP will not change only in the z-direction. This is equivalent to neglecting bending the field lines near the surface of the NP. This assumption artificially creates a small potential gradient inside each NP in the planes parallel to the flat electrode/electrolyte interface. This is, of course, incorrect since we consider metallic NPs, but this may not bring a substantial error if the radii of NPs are much larger than the Debye length, as considered in the present work. Considering for now that each particle is surrounded by some cylindrically symmetric potential distribution, $V_1(R, z)$, this potential has to be integrated over the charge density of a sphere to give

$$E = \frac{q}{2a} \int_0^a \frac{R dR}{\sqrt{a^2 - R^2}} \left[ V_1\left(R, z_0 - \sqrt{a^2 - R^2}\right) + V_1\left(R, z_0 + \sqrt{a^2 - R^2}\right) \right], \tag{20}$$

where $q$ is the total charge on the particle.

Based on the model assumed for $V_1$, the mathematical formula for $R < a$ is:

$$V_1(R, z) = \begin{cases} V(z), & z \in \left[0, z_0 - \sqrt{a^2 - R^2}\right) \\ V\left(z_0 - \sqrt{a^2 - R^2}\right), & z \in \left[z_0 - \sqrt{a^2 - R^2}, z_0 + \sqrt{a^2 - R^2}\right] \\ V\left(z - 2\sqrt{a^2 - R^2}\right), & z \in \left(z_0 + \sqrt{a^2 - R^2}, \infty\right] \end{cases} \tag{21}$$

where $V(z)$ is the Gouy-Chapman potential.

Substituting this expression into eq. (20) leads to a relatively simple final formula for this interaction:

$$E = \frac{2NkT}{a} \int_0^a \ln \frac{\coth \frac{eV_0}{4kT} + e^{-k_1(z_0-t)}}{\coth \frac{eV_0}{4kT} - e^{-k_1(z_0-t)}} \, dt, \tag{22}$$

where N is the number of charges on the particle.

### 2.1.3 Finding the effective charge on the particles, $N_{eff}$

For a 1:1 electrolyte, the Poisson-Boltzmann equation is:

$$\nabla^2 \frac{e\phi}{kT} = k^2 \sinh \frac{e\phi}{kT} \tag{23}$$

It is well known that when $\phi \ll \frac{kT}{e}$, the equation can be linearized, but not outside of this regime. The spherically symmetric solution of the linear Poisson-Boltzmann equation is

$$\phi(r) = \frac{N_{eff} e}{4\pi\epsilon_0 \epsilon} \frac{e^{ka}}{1+ka} \frac{e^{-kr}}{r}, \tag{24}$$

where $r$ is the distance from the centre of the particle, and $a$ is the particle radius.

However, the potential can also be defined as the sum of two linear screening terms, but with different screening lengths.

$$\phi(r) = \frac{N_{eff} e}{4\pi\epsilon_0 \epsilon} \frac{e^{ka}}{1+ka} \frac{e^{-kr}}{r} + \frac{(N - N_{eff}) e}{4\pi\epsilon_0 \epsilon} \frac{e^{\beta ka}}{1+\beta ka} \frac{e^{-\beta kr}}{r}, \tag{25}$$

where $\beta$ is a constant linked directly to the new screening length, which is yet to be determined.

The nonlinear Poisson-Boltzmann equation was solved, at this stage, numerically using the finite element method in COMOL Multiphysics. Eq. (25) was fitted with extremely good results to COMSOL simulations. One could calculate, based on acidity constants, pH and ligand sizes, that the total charge on each NP is about $-870e$. Fig. (3) shows, according to COMSOL, how much of this charge actually contributes to linear screening, concluding that the number is about $-300e$. Therefore, all energies that depend on the total charge are plotted for $N_{eff}$ instead, as it plays the role of a renormalized charge.

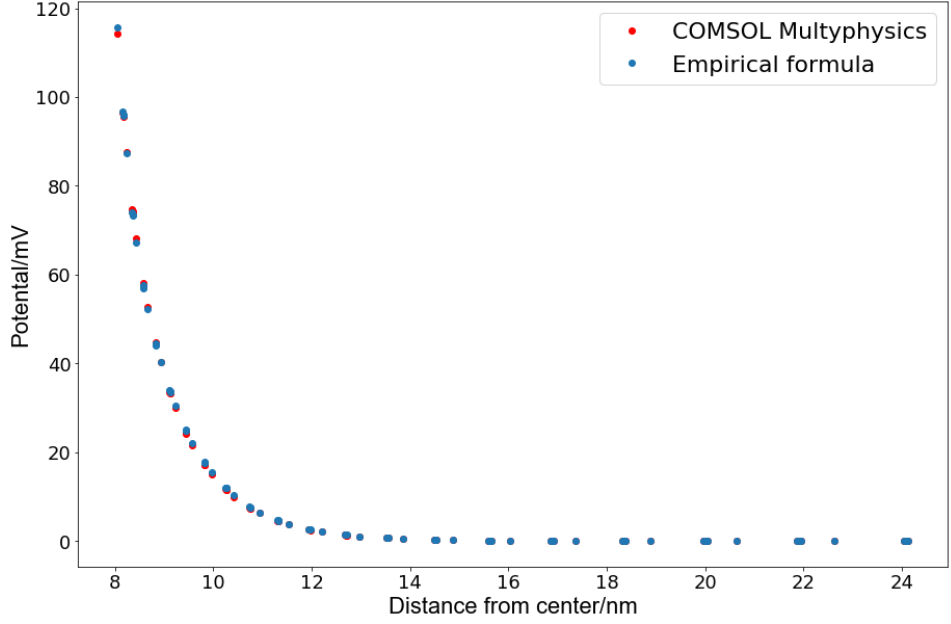

Fig. 3 Electrostatic potential of one NP in the bulk, compared to the empirical formula for potential, with $N_{eff} = 304, \beta = 5.767, N = 870$, $\epsilon = 79$, a=8 nm, k=0.805 nm$^{-1}$

### 2.1.4 Pair interaction (between two adjacent particles)

The potential of a spherical particle in eq. (8) contains, as mentioned, two terms, only this time both are needed in order to calculate the interaction energy. The first term can be calculated easily in spherical coordinates, leading to

$$\phi_0(r) = \frac{q_1}{4\pi\epsilon_0\epsilon_1} \frac{e^{-k_1 r}}{r} \frac{\sinh(k_1 a)}{k_1 a}, \tag{26}$$

where $r$ is the distance from the centre of the particle, $r > a$.

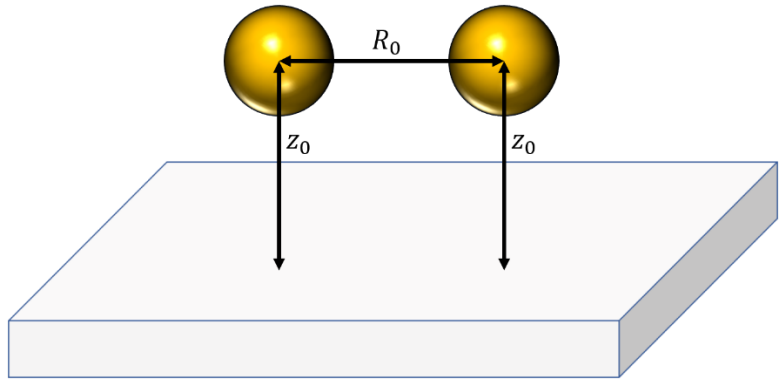

Fig.4 Pair of NPs separated by a distance $R_0$, at a distane $z_0$ from the electrode.

Starting the calculations from the linear version of the Poisson-Boltzmann equation also allows the potential generated by two particles to be written as the sum of individual potentials. Because the model aims at describing a two-dimensional NP array, this interaction will be calculated for two particles located at the same distance from the interface ($z_0$ surface-to-centre) and separated by a centre-to-centre distance $R_0$, as in fig. 4. In this case, the pair interaction energy is given by

$$W = \int d^3\vec{r}\phi_0(\vec{r})\rho(\vec{r}-\vec{R}_0) + \int d^3\vec{r}\delta\phi(\vec{r})\rho(\vec{r}-\vec{R}_0), \quad (27)$$

Which gives, after a series of manipulations, a closed form expression:

$$\boxed{\frac{W}{kT} = N^2 L_B \left(\frac{sinh(k_1 a)}{k_1 a}\right)^2 \left(\frac{e^{-k_1 R_0}}{R_0} + \int_0^\infty \frac{KdK}{\sqrt{K^2+k_1^2}} \frac{\epsilon_1\sqrt{K^2+k_1^2} - \epsilon_2\sqrt{K^2+k_2^2}}{\epsilon_1\sqrt{K^2+k_1^2} + \epsilon_2\sqrt{K^2+k_2^2}} e^{-2\sqrt{K^2+k_1^2}z_0} J_0(KR_0)\right)} \quad (28)$$

Its first term represents the interaction energy, within the linear approximation, in the absence of any dielectric interface (or in the bulk, far away from the interface). The second term is a correction caused by the presence of the interface. For an 'ideal' metal (with no static electric field penetration into it), this term is always negative, so inter-particle repulsions are weakened. However, the electrolyte concentration has a much stronger effect on the pair interaction, as one can see in figs. 5 a) and b).

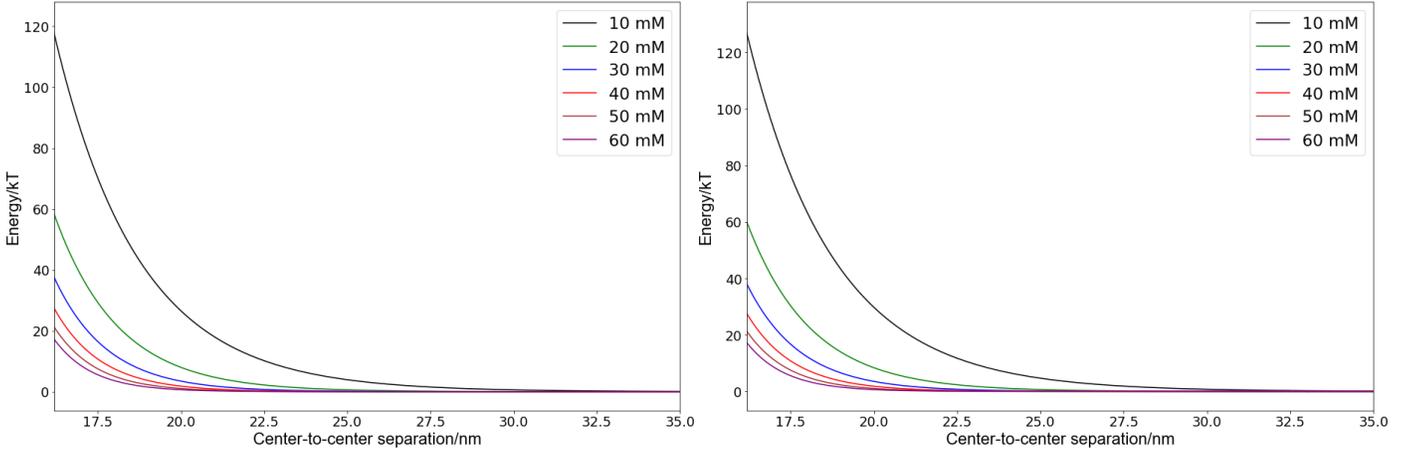

Fig. 5. Pair electrostatic interaction energy of two 8 nm-radius NPs, charged with N=-300 elementary charges, in aqueous electrolyte ($\epsilon_1 = 79$), positioned at a distance of 8.5 nm (center to surface) from a a) gold plate ($\epsilon_2 = 5.92$, $k_2 = 20$ nm$^{-1}$) and b) a TiN plate ($\epsilon_2 = 2.75$, $k_2 = 7.1$ nm$^{-1}$), respectively, as a function of electrolyte concentration. The insets show the colour coding for indicated electrolyte concentrations.

## 2.2 Van der Waals interactions

Although tunability is achieved by manipulating electrostatic interactions, van der Waals forces are always present, attracting identical NPs to each other and to the electrode, and electrostatic interactions must be able to compete with them, making the system 'electrovariable'. Strictly speaking, the problem of van der Waals interactions cannot be split into separate contributions (as we did for the linear PB equation in electrostatics). But a multi-body van der Waals equation would require a very complicated theory[32]. In the linear response approximation, a result for two point like molecules near a metallic surface, which renormalizes the electrostatic Green's function, is known[25]. The result was obtained under the assumption that all characteristic electronic excitations in the substrate metal are much faster than those in the molecules (ideal metal approximation). It shows that the presence of such a substrate diminishes the Heitler-London dispersion forces by a factor of 2/3, if the separation of the molecules from the surface is much smaller than the separation between them. Extension of this result on the case of Van der Waals interaction of NPs and with the account for frequency dependence of the dielectric function of the metal remains to be performed, but a similar kind of reduction is expected to take place there. In this manuscript we deliberately considered the maximal possible effect of Van der Waals interaction by considering their van der Waals interaction as in the bulk liquid. Mathematically, the chosen approach was the Hamaker-Lifshitz model, which calculates the interaction by summing the interaction energies of induced dipoles. The convenience of this model comes from the fact that van der Waals potential energies are written as a product between an energy constant, calculated from the frequency dependent dielectric response functions, and a geometric factor, which accounts for the shape and separation of the interacting objects.[24] Although formulas are more complicated than those for interactions between atoms, they did not pose any technical problems.

### 2.2.1 NP-electrode attraction

Geometrically, the configuration of one NP near the electrode is modelled simply as a sphere interacting with a flat surface. The Hamaker constant depends on the dielectric constants of the materials, Au for NPs, and for the electrode the material is usually a metal (Au or Ag) or a semi-metallic material (i.e. TiN or ITO). In this case, the formula for potential energy is[24]:

$$W = -\frac{A_{el.material-gold}}{6}\left(\frac{a}{z_0 - a} + \frac{a}{z_0 + a} + \ln\frac{z_0 - a}{z_0 + a}\right), \tag{29}$$

Where $A_{el.material/gold}$ is the Hamaker constant for interaction between the electrode material and gold. In order to calculate this constant, it is necessary to represent the frequency dependent dielectric constants accurately. The most convenient way of representing them is through a Drude-Lorentz formula, with two Lorenzians[16].

$$\epsilon(\omega) = \epsilon_\infty - \frac{\omega_p^2}{\omega^2 + i\gamma_p\omega} - f_1\frac{\omega_1^2}{\omega^2 - \omega_1^2 - i\gamma_1\omega} - f_2\frac{\omega_2^2}{\omega^2 - \omega_2^2 - i\gamma_2\omega}, \tag{30}$$

Here $\omega_p$ and $\gamma_p$ are the plasma frequency and plasma damping factor, $f_i$ are oscillator strengths for interband transitions, $\omega_i$ are resonance frequencies and $\gamma_i$ are damping factors for their respective interband transitions.

It is especially important for Au and TiN to reproduce interband transitions accurately, because one of them occurs in the visible range. Although experimental data for the refractive index and extinction coefficient of Au are widely available[33], finding the right Drude-Lorentz fitting parameters for TiN proved to be difficult as its optical properties are highly dependent on the Ti:N ratio.[34] Hence, even a slight difference in the deposition method of the film can lead to different reflectance spectra. The simple solution, in the end, was to find fitting parameters that accurately reproduce the reflectance of the interface in the visible region for the sample used in experiments[17].

**Table 1** Drude-Lorentz fitting parameters for Au and TiN

|     | $\epsilon_\infty$ | $\omega_p$/eV | $\gamma_p$/eV | $f_1$ | $\omega_1$/eV | $\gamma_1$/eV | $f_2$ | $\omega_2$/eV | $\gamma_2$/eV |
|-----|---------|---------|---------|---------|---------|---------|---------|---------|---------|
| Au  | 5.08961 | 9.0271  | 0.07595 | 1.42876 | 2.95297 | 0.95409 | 1.84651 | 4.06162 | 1.56389 |
| TiN | 1.16671 | 4.9652  | 3.05744 | 2.48806 | 12.87681| 22.40653| 4.84377 | 5.83046 | 5.2834  |

Table 1, however, shows Drude-Lorentz parameters for bulk Au and TiN. For gold NPs, it is important to correct the model, to account for their finite size. In this case, a difference in dielectric constant arises from the fact that the mean free path of electrons in gold is much larger than the size of a particle. However, this problem can be solved simply by taking the contribution of electron surface scattering[35]. The correction for a spherical particle is then given by:

$$\gamma_p = \gamma_p^{(0)} + \frac{3}{4} A \frac{v_F}{R}, \quad (31)$$

where $\gamma_p^{(0)}$ is the plasma damping factor of the bulk material, $A \approx 0.25$ is a constant determined experimentally, $v_F$ is Fermi velocity of electrons in gold, and $R$ is the particle radius.

The purpose behind modelling dielectric constants is to be able to sum over the entire frequency range to calculate the Hamaker constants by summing over the Matsubara frequencies $\omega = 2\pi$ (kT/$\hbar$) n:

$$A_{el.material-gold} = \frac{3kT}{2} {\sum_{n=0}^{\infty}}' \frac{\epsilon_{gold}(n) - \epsilon_{water}(n)}{\epsilon_{gold}(n) + \epsilon_{water}(n)} \frac{\epsilon_{el.material}(n) - \epsilon_{water}(n)}{\epsilon_{el.material}(n) + \epsilon_{water}(n)} \quad (32)$$

where the prime indicates that the n=0 term is weighted by 1/2. Results of the Hamaker model are shown for a metal electrode (Au) and for TiN in figs. 6 a) and b).

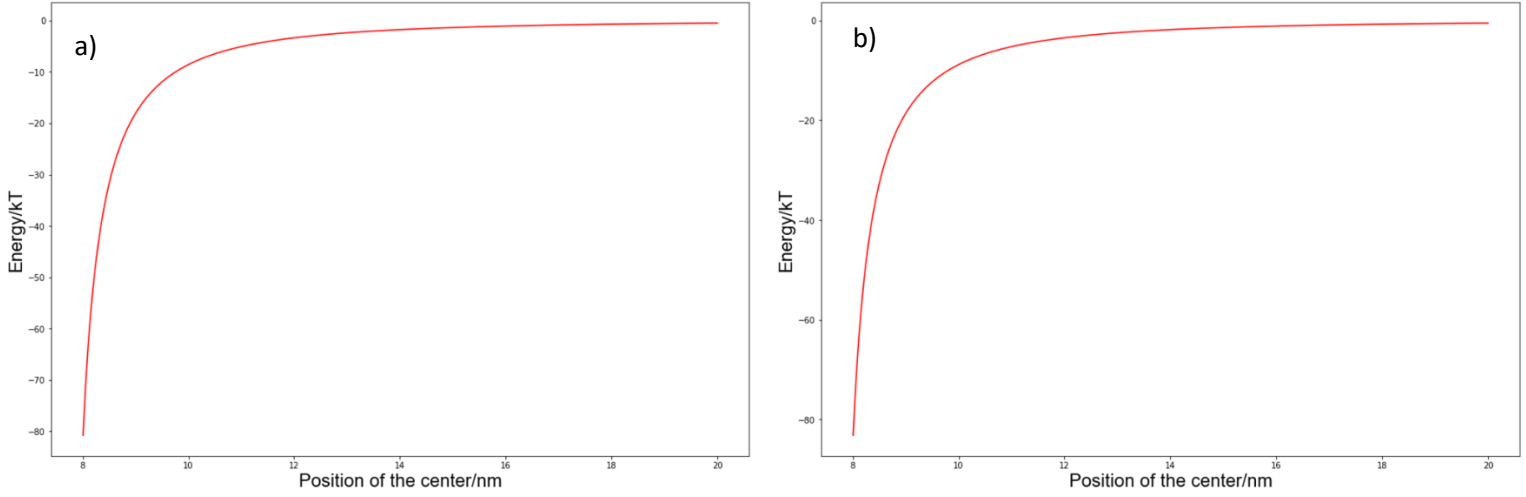

Fig. 6 Van der Waals attraction between one 8 nm-radius particle and a gold (a)/TiN (b) plate, where dielectric constants were evaluated from eq. with Drude-Lorentz parameters from table 1.

### 2.2.2 Pair interaction

The strategy for calculating attraction between two gold NPs is similar to what was presented in the previous subsection, in the sense that the Hamaker constant is calculated in the same way. The only difference is that both particles are made of gold, so eq. (31) has to be updated to:

$$A_{gold-gold} = \frac{3kT}{2} \sum_{n=0}^{\infty} \left( \frac{\epsilon_{gold}(n) - \epsilon_{water}(n)}{\epsilon_{gold}(n) + \epsilon_{water}(n)} \right)^2 \quad (33)$$

where again the n=0 term is weighted by 1/2. Nevertheless, the geometric factor of the interaction has to be changed. Because, this time, the second object is also spherical and is of the same radius, a different version of eq. (32) is required.

$$W = -\frac{A_{gold-gold}}{3} \left[ \frac{a^2}{R_0^2 - 4a^2} + \frac{a^2}{R_0^2} + \frac{1}{2} \ln\left( 1 - \frac{4a^2}{R_0^2} \right) \right] \quad (34)$$

Applying the formula above gives a van der Waals pair interaction that is, of course, independent of any distance from the interface, according to the aforementioned simplification (in reality their proximity to the conducting substrate will weaken the Van der Waals interaction).

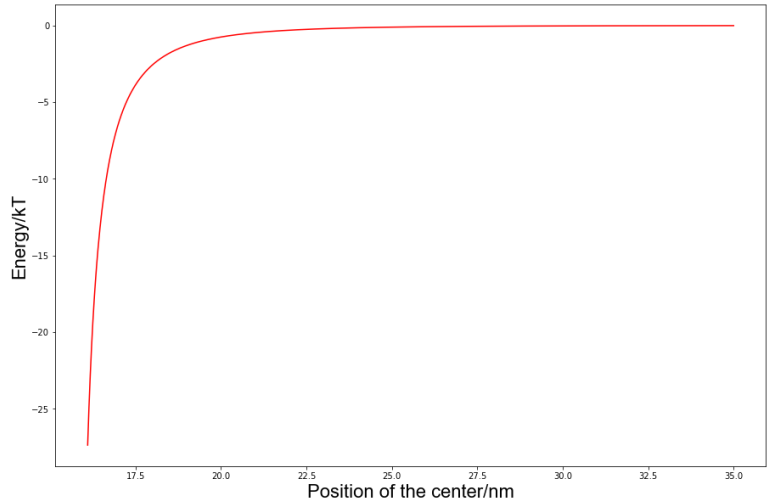

Fig. 7 Van der Waals attraction between two gold NPs, 8 nm in radius, where the gold dielectric constant was calculated with parameters from table 1.

## 3. Results and discussion

### 3.1 Net potential energy profile for NP-electrode interaction

Adding all interactions of one particle with the interface reveals an expected trend concerning electrosorption of a single particle. Even though energy values might not be accurate, fig. (8) suggests there is enough freedom to vary electrode potentials around the potential of zero charge in order to switch the overall force on each NP from attractive to repulsive.

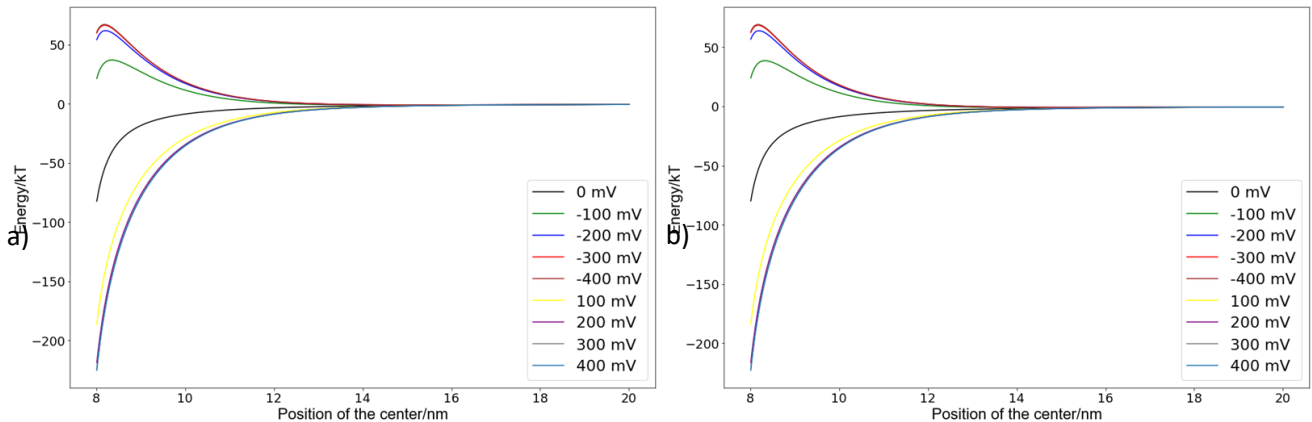

Fig. 8. Total interaction energy between one NP and the interface as a function of applied voltage, calculated by adding the van der Waals contribution (as presented in fig. 6) to the electrostatic contribution, for a gold (a) and TiN (b) substrate. The electrostatic parameters corresponding to the above curves are $\epsilon_2 = 5.92, k_2 = 20$ nm$^{-1}$ for gold and $\epsilon_2 = 2.75, k_2 = 7.1$ for TiN. The NP, 8 nm in radius, is charged with -300 e, and immersed in a 60 mM electrolyte, with $\epsilon_1 = 79$ and $k_1 = 0.805$ nm$^{-1}$.

## 3.2  Net pair interaction energy

Similarly, all contributions to the pair interaction are also combined. The concentration effect is clear from fig. (9), the higher the concentration the lower the repulsion. However, even for the largest concentration on the graph, repulsion is still strong enough to win over van der Waals attraction and stop the particles from agglomerating into clusters.

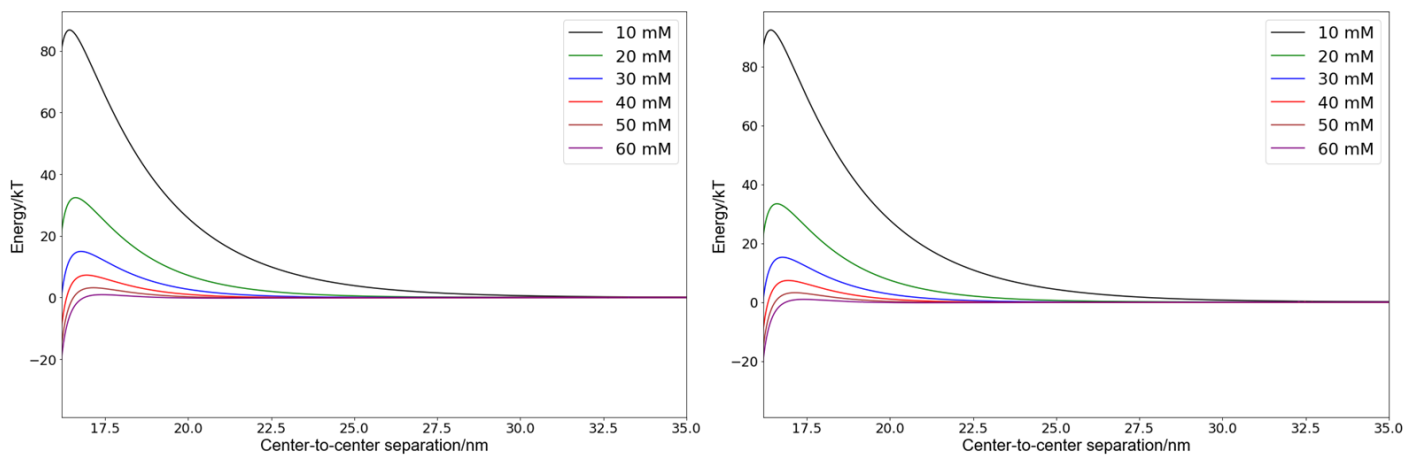

Fig. 9 a) Total pair interaction energy, as a function of electrolyte concentration, calculated by adding the van der Waals contribution (as presented in fig. 7) to the electrostatic contribution shown in fig. 5.

## 3.3  Mean-field electrosorption isotherm

After developing a consistent model for the key interactions in the system (both van der Waals and electrostatic), one may ask a question, how do these forces affect together the NP assembly at the surface, when it becomes favourable? To answer it, the stability of the NP array was analysed using the Ising model[22].

Originally, the Ising model was meant to deal with interacting magnetic domains arranged in a lattice, which are characterised by spins, and the interaction of these domains with an external magnetic field. Therefore, key interaction parameters occur: $J_{ij}$, the pair interaction energies between two spins ($i$ and $j$), and $h_i$, the interaction of each spin with the external field. These parameters and, more importantly, their balance, determine the behaviour of the lattice and whether a phase transition occurs (from paramagnetism to ferromagnetism or vice versa). Hence, the macrostate of the lattice depends on the microstates of individual spins and their interaction parameters, and it is described by the following hamiltonian function

$$H = - \sum_{pairs<i,j>} J_{ij} S_i S_j + \sum_i h_i S_i, \quad (35)$$

where $S_i$ are the spins and can take only two orientations, up or down, in an external field $h$ : $S_i = \{\pm 1\}$.

Later, the Ising model was successfully applied to adsorption, and we will use a similar approach by mapping on it the problem of electrosorption of NPs. The first assumption that we will make is that NPs, when adsorbed at the interface, arrange themselves into a hexagonal lattice, so all particles will be equally spaced. Next, because electrostatic interactions are exponentially screened, whereas Van der Waals interactions decay no slower than inverse cube of the distance between NPs which is short range in two dimensions, we can safely take into account only the nearest neighbour interactions. This collapses all pair interaction parameters into one, $J$, containing both electrostatic and van der Waals contributions at distances between NP corresponding to their dense packing. The fact that practically NPs will settle, if adsorbed, at some distances from each other, will be taken into account through the value of the *coverage*, $\theta$ – the probability that a site of that hexagonal lattice is occupied, 0<$\theta$<1: the more sparsely the NPs settle at the interface, the lower $\theta$ will be. Because the interface interacts in the same way with all particles (within the linear approximation), the 'external field' will also have only one value, $h$. The parameter, $h$, will have a meaning of the interaction energy between a NP and the interface; it will contain the contributions from van der Waals attraction, image force, and interaction with the charge on the electrode.

Thus, each site of this isotropic NP lattice (with $N$ being the total number of sites) can be either occupied or unoccupied. It is further convenient to move from the spin variables, $S_i$, to $B_i$, defined as the occupation of lattice site $i$. Its values are, in this case, $B_i \in \{0,1\}$. Updating eq. (35) with these changes gives:

$$H = J \sum_{i=1}^{N} \sum_{j \in neighbours(i)} B_i B_j + h \sum_{i=1}^{N} B_i \tag{36}$$

One can easily see from eq. (35) that, for two adjacent lattice sites, an interaction occurs only if they are both occupied, with $J > 0$ describing the strength of repulsion. Similarly, the interface only interacts with an occupied lattice site.

The only analytical solution for the two-dimensional Ising model is the one derived by Onsager, for a square lattice and in the absence of external fields. For our estimates it would be sufficient, however, to use the simpler, mean-field approximation. In other words, each site is assumed to interact with the average occupancy of the entire lattice, so the Hamiltonian becomes

$$H = (zJ\langle B \rangle + h) \sum_{i=1}^{N} B_i \tag{37}$$

The final step in the Ising model calculation is to find the average value of the occupancy, which also represents the coverage of the surface (as a fraction of the number of lattice sites) for a given lattice constant. Considering that occupancies obey the Boltzmann distribution, $\langle B \rangle$ can be written as

$$\langle B \rangle = \frac{e^{-\frac{(zJ\langle B \rangle + h)}{kT}}}{e^{-\frac{(zJ\langle B \rangle + h)}{kT}} + 1} \tag{38}$$

This well-known equation does not have an analytical solution for $\langle B \rangle$ as a function of $h$, but there is one for $h$ as a function of $\langle B \rangle$:

$$\frac{h}{kT} = \ln\frac{1 - \langle B \rangle}{\langle B \rangle} - Z\frac{J}{kT}\langle B \rangle \tag{39}$$

For given values of $J/kT$, plotting $h/kT$ vs $\langle B \rangle$ in the interval between 0 and 1, and rotating the coordinate system by 90 degrees, one obtains a graph of the coverage $\langle B \rangle$ as a function of $h$. Of course, the value of the lattice constant needs to be set in order to evaluate $J$. The chosen value corresponds to a relatively dense lattice, where NP surfaces are 2 nm apart. The resulting computations are shown in figs 16 for metallic and semi-metallic electrodes, in which the values of $J$ have been calculated using Eqs (28) and (34), and $h$ related to voltage subject to Eqs. (17), (22) and (29), each equation representing a separate contribution.

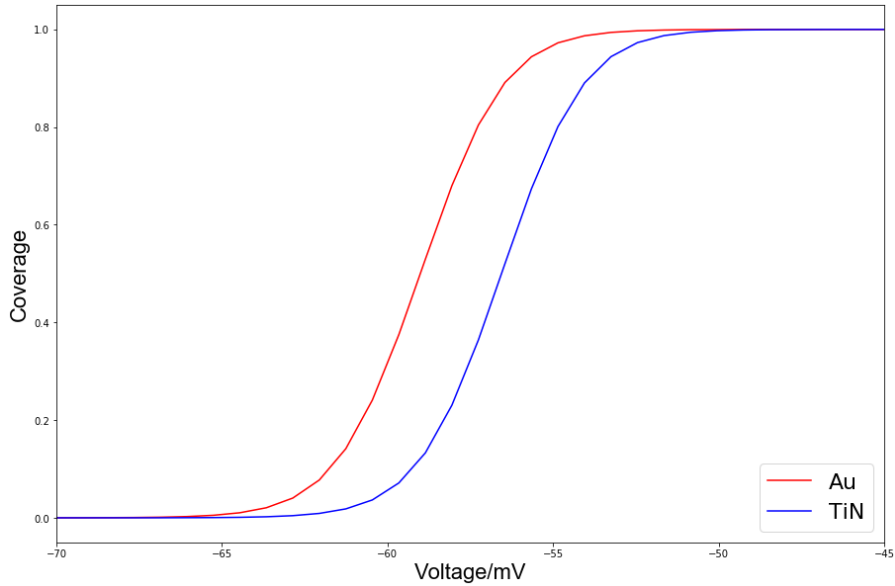

Fig. 10 Electrosorption isotherm: lattice coverage as a function of applied potential, for Au and TiN substrates, based on the interaction energies shown in figs. 8 and 9, evaluated for a lattice constant of 19 nm and particle centers at 8.5 nm from the interface.

For both types of electrode, theory predicts a very narrow voltage interval (about 30 mV), where the assembly of each lattice changes from unfavourable to favourable. Apart from the conceptual shortcomings of the oversimplified theory presented above, it is also possible that the voltage window for such a transition, in the significantly larger range seen in experiments[17] may come from different sources – some surface roughness, dispersion in particle sizes, inhomogeneity of lattices, multiple reasons. The neglect of nonlinear effects can generate exaggerated interaction energy values, while in practice, energies could be smaller, entropy widening the voltage interval where assembly takes place.

Theoretical predictions on the density of NP arrays are given in the form of electrosorption isotherms. As shown in section 3.3, changing the potential drop across the interface can easily shift the balance between pair interaction and NP-electrode interaction, allowing particles to assemble or disassemble. However, the simplicity of the theoretical framework is likely to give inaccurate numerical results, while giving a good qualitative picture. In order to solve this problem or, at least, improve the estimates, one has to return to the assumptions behind electrostatic forces. Regarded as an important correction would be the fact that NPs are polarisable. So, the pair interaction, for example, has to take into account that each charged particle polarises the other particles around it. This problem was solved to some degree[36], (within linear Poisson-Boltzmann regime) but the problem of nonlinear effects still remains. Equally on the Van der Waals front, it would be better to take into account the effect of the substrate on pair interactions. But this may be of lower importance, because, as we see, the Van der Waals interactions are generally substantially smaller than the electrostatic ones, unless we screen the latter stronger, by electrolyte concentrations higher than those considered in this study.

## 4. Conclusion

The analysis presented in this study highlights the following effects influencing assembly or disassembly of NP arrays at electrochemical solid-liquid interfaces:

1. interaction energies of NPs with the interface are highly dependent on applied voltage

2. van der Waals and image forces tend to combine into an overall attractive force

3. the pair electrostatic interaction energy does not depend on the applied voltage, but is largely influenced by electrolyte concentration through screening. Had we however taken into account the increase of concentration of counterions of the same sign as of NPs (considered to be negative in this work), we would have stronger repulsion between them for positive electrode polarisations and a weaker repulsion for negative ones, which would have smoothened the coverage dependence on electrode potential, increasing the voltage range where the crossover between assembly and disassembly would take place.

But all in all, we have shown that the reversible assembly at the solid-liquid interface is made possible simply by changing the voltage applied across the interface.

# Appendix

## 1. Electrostatics

### 1. Charge density of a spherical shell of radius a centered at $z_0$

$$\rho(R,z) = C\delta\left(\sqrt{R^2 + (z-z_0)^2} - a\right) (A1)$$

$$\iiint \rho(R,z)dV = q$$

$$C\int_{-\infty}^{\infty}\int_0^{2\pi}\int_0^{\infty} \delta\left(\sqrt{R^2 + (z-z_0)^2} - a\right) RdRd\theta dz = q$$

$$2\pi C\int_{-\infty}^{\infty} dz \int_0^{\infty} \delta\left(\sqrt{R^2 + (z-z_0)^2} - a\right) RdR = q$$

If $(z-z_0)^2 > a^2$, then $\left(\sqrt{R^2 + (z-z_0)^2} - a\right) > 0$, so $\delta\left(\sqrt{R^2 + (z-z_0)^2} - a\right) = 0$. The only way the integral is not equal to 0 is when $(z-z_0)^2 < a^2$, which is equivalent to $z \in [z_0 - a, z_0 + a]$.

When $\sqrt{R^2 + (z-z_0)^2} = a$, $R = \sqrt{a^2 - (z-z_0)^2}$ because R has to be positive. Therefore, the integral becomes

$$2\pi C\int_{z_0-a}^{z_0+a} dz \int_0^{\infty} \frac{\delta\left(R - \sqrt{a^2 - (z-z_0)^2}\right)}{\left|\frac{R}{\sqrt{R^2 - (z-z_0)^2}}\right|_{R=\sqrt{a^2-(z-z_0)^2}}} RdR = q$$

$$2\pi C\int_{z_0-a}^{z_0+a} dz \int_0^{\infty} \frac{\delta\left(R - \sqrt{a^2 - (z-z_0)^2}\right)}{\frac{\sqrt{a^2 - (z-z_0)^2}}{a}} RdR = q$$

$$2\pi aC\int_{z_0-a}^{z_0+a} dz = q$$

$$C = \frac{q}{4\pi a^2}$$

So the formula for charge density is:

$$\boxed{\rho(R,z) = \frac{q}{4\pi a^2}\delta\left(\sqrt{R^2 + (z-z_0)^2} - a\right)} (A2)$$

### 2. Fourier transform of the charge density

$$\tilde{\rho}(K,z) = \frac{1}{(2\pi)^2}\int_0^\infty \int_{-\pi}^\pi e^{-iKR\cos\theta}\rho(R,z)Rd\theta dR$$

$$\boxed{\tilde{\rho}(K,z) = \frac{1}{2\pi}\int_0^\infty \rho(R,z)J_0(kR)RdR}$$

$$\tilde{\rho}(K,z) = \frac{1}{2\pi}\int_0^\infty J_0(KR)\frac{q}{4\pi a^2}\delta\left(\sqrt{R^2+(z-z_0)^2}-a\right)RdR$$

$$\tilde{\rho}(K,z) = \frac{q}{8\pi^2 a^2}\int_0^\infty J_0(KR)\frac{\delta\left(R-\sqrt{a^2-(z-z_0)^2}\right)}{\frac{\sqrt{a^2-(z-z_0)^2}}{a}}\Theta(a-|z-z_0|)RdR$$

$$\tilde{\rho}(K,z) = \frac{q}{8\pi^2 a}\frac{\Theta(a-|z-z_0|)}{\sqrt{a^2-(z-z_0)^2}}\int_0^\infty J_0(KR)\delta\left(R-\sqrt{a^2-(z-z_0)^2}\right)RdR$$

$$\boxed{\tilde{\rho}(K,z) = \frac{q}{8\pi^2 a}J_0\left(K\sqrt{a^2-(z-z_0)^2}\right)\Theta(a-|z-z_0|)} \quad (A3)$$

**Checking the limiting case $a \to 0$ (point charge):**

$$\lim_{a\to 0}\tilde{\rho}(K,z) = \frac{q}{8\pi^2}\lim_{a\to 0}\frac{J_0(\cdots}{}$$

$$\frac{q}{8\pi^2}\lim_{a\to 0}-\frac{Ka}{\sqrt{a^2-(z-z_0)^2}}J_1\left(K\sqrt{a^2-(z-z_0)^2}\right)\Theta(a-|z-z_0|) + \frac{q}{8\pi^2}\lim_{a\to 0}J_0\left(K\sqrt{a^2-(z-z_0)^2}\right)\delta(a-|\cdots$$

### 1.3 Finding the electrostatic potential as solution of the Poisson-Boltzmann equation

$$\nabla^2\phi_1 - k_1^2\phi_1 = -\frac{\rho}{\epsilon_0\epsilon_1} \quad (A4)$$

$$\nabla^2\phi_2 - k_2^2\phi_2 = 0 \quad (A5)$$

The solutions of these equations have to be subjected to the following set of boundary conditions, representing continuity of potential and continuity of the normal component of electric induction at the interface.

$$\phi_1(\vec{R},0) = \phi_2(\vec{R},0) \quad (A6)$$

$$\epsilon_1\frac{\partial\phi_1}{\partial z}\bigg|_{z=0} = \epsilon_2\frac{\partial\phi_2}{\partial z}\bigg|_{z=0} \quad (A7)$$

**1.3.1 Writing $\phi_1$ and $\phi_2$ in terms of their Fourier transforms:**

$$\phi_1(\vec{R},z) = \int e^{i\vec{K}\cdot\vec{R}}\tilde{\phi}_1(\vec{K},z)d\vec{K}$$

$$\phi_2(\vec{R},z) = \int e^{i\vec{K}\cdot\vec{R}} \tilde{\phi}_2(\vec{K},z) d\vec{K}$$

Substituting the expressions above into the equations leads to:

$$\int e^{i\vec{K}\cdot\vec{R}} \left( -K^2 \tilde{\phi}_1 + \frac{\partial^2 \tilde{\phi}_1}{\partial z^2} \right) d\vec{K} = k_1^2 \int e^{i\vec{K}\cdot\vec{R}} \tilde{\phi}_1 - \frac{1}{\epsilon_0 \epsilon_1} \int e^{i\vec{K}\cdot\vec{R}} \tilde{\rho} d\vec{K}$$

$$\int e^{i\vec{K}\cdot\vec{R}} \left( \frac{\partial^2 \tilde{\phi}_1}{dz^2} - (K^2 + k_1^2)\tilde{\phi}_1 + \frac{1}{\epsilon_0 \epsilon_1}\tilde{\rho} \right) d\vec{K} = 0$$

$$\frac{\partial^2 \tilde{\phi}_1}{\partial z^2} - (K^2 + k_1^2)\tilde{\phi}_1 = -\frac{\tilde{\rho}}{\epsilon_0 \epsilon_1} \quad (A8)$$

Using the same procedure for $\phi_2$ gives the equation:

$$\frac{\partial^2 \tilde{\phi}_2}{\partial z^2} - (K^2 + k_2^2)\tilde{\phi}_2 = 0 \quad (A9)$$

The boundary conditions can also be written using the Fourier transforms of these potentials:

$$\tilde{\phi}_1(\vec{K},0) = \tilde{\phi}_2(\vec{K},0) \quad (A10)$$

$$\epsilon_1 \frac{\partial \tilde{\phi}_1}{\partial z}\bigg|_{z=0} = \epsilon_2 \frac{\partial \tilde{\phi}_2}{\partial z}\bigg|_{z=0} \quad (A11)$$

Equation (A9) gives the potential in medium 2. The solution of this equation is known to be:

$$\tilde{\phi}_2(K,z) = A e^{\sqrt{K^2+k_2^2}\,z} + B e^{-\sqrt{K^2+k_2^2}\,z} \quad (A12)$$

If we go infinitely far away from any charge distribution in medium 2 ($z \to -\infty$) the potential has to approach 0, so $\tilde{\phi}_2$ has to approach 0 as well. Therefore, $B = 0$. So we are left with

$$\tilde{\phi}_2(K,z) = A e^{\sqrt{K^2+k_2^2}\,z} \quad (A13)$$

The next step is to find the potential in medium 1 by solving equation (A8). This can be achieved by finding the Green's function of equation (A8).

$$\frac{\partial^2 \tilde{G}}{\partial z^2} - (K^2 + k_1^2)\tilde{G} = -\delta(z - z_0) \quad (A14)$$

The most general solution of the equation above is composed of the general solution of the associated homogeneous equation and any particular solution of that equation.

$$\tilde{G}(K,z,z_0) = \tilde{G}_h(K,z,z_0) + \tilde{G}_p(K,z,z_0) \quad (A15)$$

$$\tilde{G}_h(K, z, z_0) = Ce^{\sqrt{K^2+k_1^2}\,z} + De^{-\sqrt{K^2+k_1^2}\,z} \quad (A16)$$

In order to find a particular solution, the method of Fourier transforms was used. In other words, a one-dimensional Fourier transform of $\tilde{G}_p(K,z)$ is defined and calculated, then the solution is recovered by taking the inverse Fourier transform.

$$\tilde{G}_F(K,w) = \frac{1}{2\pi}\int_{-\infty}^{\infty} e^{-izw}\tilde{G}_p(K,z)dz$$

$$\tilde{G}_p(K,z,z_0) = \int_{-\infty}^{\infty} e^{izw}\tilde{G}_F(K,w)dw$$

$$\int_{-\infty}^{\infty} e^{izw}\left(-w^2\tilde{G}_F\right)dw - \int_{-\infty}^{\infty} e^{izw}(K^2+k_1^2)\tilde{G}_F dw = -\frac{1}{2\pi}\int_{-\infty}^{\infty} e^{i(z-z_0)w}dw$$

$$\int_{-\infty}^{\infty} e^{izw}\left((w^2+K^2+k_1^2)\tilde{G}_F - \frac{e^{-iz_0 w}}{2\pi}\right)dw = 0$$

$$\tilde{G}_F = \frac{e^{-iz_0 w}}{2\pi}\frac{1}{w^2+K^2+k_1^2}$$

$$\tilde{G}_p(K,z,z_0) = \frac{1}{2\pi}\int_{-\infty}^{\infty}\frac{e^{i(z-z_0)w}}{w^2+K^2+k_1^2}dw$$

This integral can be easily evaluated using contour integration on the upper and lower halves of a circle of infinite radius, centered at the origin.

$$\tilde{G}_p(K,z,z_0) = \frac{1}{2}\frac{e^{-\sqrt{K^2+k_1^2}\,|z-z_0|}}{\sqrt{K^2+k_1^2}} \quad (A17)$$

So, Green's function is simply

$$\tilde{G}(K,z,z_0) = Ce^{\sqrt{K^2+k_1^2}\,z} + De^{-\sqrt{K^2+k_1^2}\,z} + \frac{1}{2}\frac{e^{-\sqrt{K^2+k_1^2}\,|z-z_0|}}{\sqrt{K^2+k_1^2}} \quad (A18)$$

But a physically meaningful potential has to vanish when $z \to \infty$, so the Green's function leading to that potential also has to vanish at infinity.

$$\boxed{\tilde{G}(K,z,z_0) = De^{-\sqrt{K^2+k_1^2}\,z} + \frac{1}{2}\frac{e^{-\sqrt{K^2+k_1^2}\,|z-z_0|}}{\sqrt{K^2+k_1^2}}} \quad (A19)$$

In this case, the can be expressed in terms of the Green's function as follows:

$$\tilde{\phi}_1\left(\vec{K},z\right) = \frac{1}{\epsilon_0\epsilon_1}\int_0^{\infty}\tilde{G}(K,z,z_0)\tilde{\rho}\left(\vec{K},z_0\right)dz_0 \quad (A20)$$

$$\tilde{\phi}_1\left(\vec{K},z\right) = \frac{1}{2\epsilon_0\epsilon_1}\int_0^\infty \frac{e^{-\sqrt{K^2+k_1^2}|z-z_0|}}{\sqrt{K^2+k_1^2}}\tilde{\rho}\left(\vec{K},z_0\right)dz_0 + \frac{D}{\epsilon_0\epsilon_1}e^{-\sqrt{K^2+k_1^2}z}\int_0^\infty \tilde{\rho}\left(\vec{K},z_0\right)dz_0 \quad (A21)$$

### 1.3.2 Applying the boundary conditions:

$$\tilde{\phi}_1\left(\vec{K},0\right) = \frac{1}{2\epsilon_0\epsilon_1}\int_0^\infty \frac{e^{-\sqrt{K^2+k_1^2}z_0}}{\sqrt{K^2+k_1^2}}\tilde{\rho}\left(\vec{K},z_0\right)dz_0 + \frac{D}{\epsilon_0\epsilon_1}\int_0^\infty \tilde{\rho}\left(\vec{K},z_0\right)dz_0 \quad (A22)$$

$$\frac{\partial \tilde{\phi}_1}{\partial z} = -\frac{1}{2\epsilon_0\epsilon_1}\int_0^\infty e^{-\sqrt{K^2+k_1^2}|z-z_0|}\operatorname{sgn}(z-z_0)\tilde{\rho}\left(\vec{K},z_0\right)dz_0 - \frac{D\sqrt{K^2+k_1^2}}{\epsilon_0\epsilon_1}e^{-\sqrt{K^2+k_1^2}z}\int_0^\infty \tilde{\rho}\left(\vec{K},z_0\right)dz_0$$

$$\left.\frac{\partial \tilde{\phi}_1}{\partial z}\right|_{z=0} = \frac{1}{2\epsilon_0\epsilon_1}\int_0^\infty e^{-\sqrt{K^2+k_1^2}z_0}\tilde{\rho}\left(\vec{K},z_0\right)dz_0 - \frac{D\sqrt{K^2+k_1^2}}{\epsilon_0\epsilon_1}\int_0^\infty \tilde{\rho}\left(\vec{K},z_0\right)dz_0 \quad (A23)$$

$$\tilde{\phi}_2\left(\vec{K},0\right) = A \quad (A24)$$

$$\left.\frac{\partial \tilde{\phi}_2}{\partial z}\right|_{z=0} = A\sqrt{K^2+k_2^2} \quad (A25)$$

Therefore, the two equations extracted from the boundary conditions are:

$$A = \frac{1}{2\epsilon_0\epsilon_1}\int_0^\infty \frac{e^{-\sqrt{K^2+k_1^2}z_0}}{\sqrt{K^2+k_1^2}}\tilde{\rho}\left(\vec{K},z_0\right)dz_0 + \frac{D}{\epsilon_0\epsilon_1}\int_0^\infty \tilde{\rho}\left(\vec{K},z_0\right)dz_0 \quad (A26)$$

$$A\frac{\epsilon_2\sqrt{K^2+k_2^2}}{\epsilon_1\sqrt{K^2+k_1^2}} = \frac{1}{2\epsilon_0\epsilon_1}\int_0^\infty \frac{e^{-\sqrt{K^2+k_1^2}z_0}}{\sqrt{K^2+k_1^2}}\tilde{\rho}\left(\vec{K},z_0\right)dz_0 - \frac{D}{\epsilon_0\epsilon_1}\int_0^\infty \tilde{\rho}\left(\vec{K},z_0\right)dz_0 \quad (A27)$$

Adding them up leads to:

$$A\left(1 + \frac{\epsilon_2\sqrt{K^2+k_2^2}}{\epsilon_1\sqrt{K^2+k_1^2}}\right) = \frac{1}{\epsilon_0\epsilon_1\sqrt{K^2+k_1^2}}\int_0^\infty e^{-\sqrt{K^2+k_1^2}z_0}\tilde{\rho}\left(\vec{K},z_0\right)dz_0 \quad (A28)$$

$$A = \frac{\int_0^\infty e^{-\sqrt{K^2+k_1^2}z_0}\tilde{\rho}\left(\vec{K},z_0\right)dz_0}{\epsilon_0\left(\epsilon_1\sqrt{K^2+k_1^2} + \epsilon_2\sqrt{K^2+k_2^2}\right)}$$

Subtracting the two equations leads to:

$$A\left(1 - \frac{\epsilon_2\sqrt{K^2+k_2^2}}{\epsilon_1\sqrt{K^2+k_1^2}}\right) = \frac{2D}{\epsilon_0\epsilon_1}\int_0^\infty \tilde{\rho}\left(\vec{K},z_0\right)dz_0$$

$$\frac{1}{\epsilon_0 \epsilon_1 \sqrt{K^2+k_1^2}} \frac{\epsilon_1 \sqrt{K^2+k_1^2} - \epsilon_2 \sqrt{K^2+k_2^2}}{\epsilon_1 \sqrt{K^2+k_1^2} + \epsilon_2 \sqrt{K^2+k_2^2}} \int_0^\infty e^{-\sqrt{K^2+k_1^2} z_0} \tilde{\rho}\left(\vec{K}, z_0\right) dz_0 = \frac{2D}{\epsilon_0 \epsilon_1} \int_0^\infty \tilde{\rho}\left(\vec{K}, z_0\right) dz_0$$

$$D = \frac{1}{2} \frac{\epsilon_1 \sqrt{K^2+k_1^2} - \epsilon_2 \sqrt{K^2+k_2^2}}{\epsilon_1 \sqrt{K^2+k_1^2} + \epsilon_2 \sqrt{K^2+k_2^2}} \frac{\int_0^\infty \frac{e^{-\sqrt{K^2+k_1^2} z_0}}{\sqrt{K^2+k_1^2}} \tilde{\rho}\left(\vec{K}, z_0\right) dz_0}{\int_0^\infty \tilde{\rho}\left(\vec{K}, z_0\right) dz_0} \quad (A29)$$

Plugging in the constants gives the Fourier transforms of the potentials:

$$\boxed{\tilde{\phi}_1\left(\vec{K}, z\right) = \frac{1}{2\epsilon_0 \epsilon_1 \sqrt{K^2+k_1^2}} \left( \int_0^\infty e^{-\sqrt{K^2+k_1^2}|z-z_0|} \tilde{\rho}\left(\vec{K}, z_0\right) dz_0 + \frac{\epsilon_1 \sqrt{K^2+k_1^2} - \epsilon_2 \sqrt{K^2+k_2^2}}{\epsilon_1 \sqrt{K^2+k_1^2} + \epsilon_2 \sqrt{K^2+k_2^2}} \int_0^\infty e^{-\sqrt{K^2+k_1^2}(z+z_0)} \tilde{\rho}\left(\vec{K}, z_0\right) dz_0 \right)} \quad (A30)$$

$$\boxed{\tilde{\phi}_2\left(\vec{K}, z\right) = \frac{\int_0^\infty e^{\sqrt{K^2+k_2^2} z - \sqrt{K^2+k_1^2} z_0} \tilde{\rho}\left(\vec{K}, z_0\right) dz_0}{\epsilon_0 \left( \epsilon_1 \sqrt{K^2+k_1^2} + \epsilon_2 \sqrt{K^2+k_2^2} \right)}} \quad (A31)$$

In order to see how the presence of the interface influences the potential created by the charge distribution $\rho\left(\vec{R}, z\right)$ we have to first see how the potential looks like without the interface. This result can be derived directly from equation (A30) by making the interface 'disappear' from an electromagnetic viewpoint, using the following conditions: $\epsilon_1 = \epsilon_2$ and $k_1 = k_2$. Therefore, the potential without the interface is:

$$\tilde{\phi}_0\left(\vec{K}, z\right) = \frac{\int_0^\infty e^{-\sqrt{K^2+k_1^2}|z-z_0|} \tilde{\rho}\left(\vec{K}, z_0\right) dz_0}{2\epsilon_0 \epsilon_1 \sqrt{K^2+k_1^2}} \quad (A32)$$

The overall potential can be expressed as a sum of the self-potential $\tilde{\phi}_0$ and another term

$$\delta\tilde{\phi}\left(\vec{K}, z\right) = \frac{1}{2\epsilon_0 \epsilon_1 \sqrt{K^2+k_1^2}} \frac{\epsilon_1 \sqrt{K^2+k_1^2} - \epsilon_2 \sqrt{K^2+k_2^2}}{\epsilon_1 \sqrt{K^2+k_1^2} + \epsilon_2 \sqrt{K^2+k_2^2}} \int_0^\infty e^{-\sqrt{K^2+k_1^2}(z+z_0)} \tilde{\rho}\left(\vec{K}, z_0\right) dz_0 \quad (A33)$$

$$\boxed{\tilde{\phi}_1\left(\vec{K}, z\right) = \tilde{\phi}_0\left(\vec{K}, z\right) + \delta\tilde{\phi}\left(\vec{K}, z\right)} \quad (A34)$$

The next step is to evaluate the energy of this system, the total energy required to build it from point charges initially being infinitely far away from each other. The energy of our system, described by a charge density $\rho\left(\vec{R}, z\right)$, which generates a potential $\phi\left(\vec{R}, z\right)$ is given by the following expression:

$$W = \frac{1}{2}\int_0^\infty dz \int d\vec{R}\, \phi(\vec{R},z)\, \rho(\vec{R},z) \quad (A35)$$

Here the integral over $\vec{R}$ covers the entire xy plane and the integral over z is taken only between 0 and infinity because there is charge density only on one side of the interface.

In order to be able to use the found Fourier transform of the potential the energy expression has to be rearranged.

$$W = \frac{1}{2}\int_0^\infty dz \int d\vec{R}\, \rho(\vec{R},z) \int d\vec{K}\, \tilde{\phi}(\vec{K},z)\, e^{i\vec{K}\cdot\vec{R}}$$

$$W = \frac{1}{2}\int_0^\infty dz \int d\vec{K}\, \tilde{\phi}(\vec{K},z) \int d\vec{R}\, \rho(\vec{R},z)\, e^{-i(-\vec{K}\cdot\vec{R})}$$

$$\boxed{W = 2\pi^2 \int_0^\infty dz \int d\vec{K}\, \tilde{\phi}(\vec{K},z)\, \tilde{\rho}(-\vec{K},z)} \quad (A36)$$

Substituting the expression of the potential we get

$$W = 2\pi^2 \int_0^\infty dz \int d\vec{K}\, \left(\tilde{\phi}_0(\vec{K},z) + \delta\tilde{\phi}(\vec{K},z)\right) \tilde{\rho}(-\vec{K},z)$$

$$W = 2\pi^2 \int_0^\infty dz \int d\vec{K}\, \tilde{\phi}_0(\vec{K},z)\, \tilde{\rho}(-\vec{K},z) + 2\pi^2 \int_0^\infty dz \int d\vec{K}\, \delta\tilde{\phi}(\vec{K},z)\, \tilde{\rho}(-\vec{K},z) \quad (A37)$$

$$W = W_0 + \delta W \quad (A38)$$

In the expression above $W_0$ is the self energy of the charge distribution and $\delta W$ is the image potential energy of the charge distribution. They are given by the following equations:

$$W_0 = 2\pi^2 \int_0^\infty dz \int d\vec{K}\, \tilde{\phi}_0(\vec{K},z)\, \tilde{\rho}(-\vec{K},z) \quad (A39)$$

$$\delta W = 2\pi^2 \int_0^\infty dz \int d\vec{K}\, \delta\tilde{\phi}(\vec{K},z)\, \tilde{\rho}(-\vec{K},z) \quad (A40)$$

The important part of the energy is $\delta W$, because it represents the potential energy between the charge distribution and the interface. To proceed with its calculation the potential $\delta\tilde{\phi}$ has to be substituted into equation (A40).

$$\delta W = \frac{\pi^2}{\epsilon_0 \epsilon_1}\int_0^\infty dz \int \frac{d\vec{K}}{\sqrt{K^2+k_1^2}} \frac{\epsilon_1\sqrt{K^2+k_1^2} - \epsilon_2\sqrt{K^2+k_2^2}}{\epsilon_1\sqrt{K^2+k_1^2} + \epsilon_2\sqrt{K^2+k_2^2}} \tilde{\rho}(-\vec{K},z) \int_0^\infty e^{-\sqrt{K^2+k_1^2}(z+z_0)} \tilde{\rho}(\vec{K},z_0)\, dz_0$$

$$\delta W = \frac{\pi^2}{\epsilon_0 \epsilon_1}\int \frac{d\vec{K}}{\sqrt{K^2+k_1^2}} \frac{\epsilon_1\sqrt{K^2+k_1^2} - \epsilon_2\sqrt{K^2+k_2^2}}{\epsilon_1\sqrt{K^2+k_1^2} + \epsilon_2\sqrt{K^2+k_2^2}} \int_0^\infty e^{-\sqrt{K^2+k_1^2}z} \tilde{\rho}(-\vec{K},z)\, dz \int_0^\infty e^{-\sqrt{K^2+k_1^2}z'} \tilde{\rho}(\vec{K},z')\, dz'$$

If we swap z and z' the energy expression does not change. Therefore, we can define

$$\bar{\rho}\left(\vec{K},z\right) = \sqrt{\tilde{\rho}\left(-\vec{K},z\right)\tilde{\rho}\left(\vec{K},z\right)} \quad (A41)$$

$$\delta W = \frac{\pi^2}{\epsilon_0 \epsilon_1} \int \frac{d\vec{K}}{\sqrt{K^2+k_1^2}} \frac{\epsilon_1\sqrt{K^2+k_1^2} - \epsilon_2\sqrt{K^2+k_2^2}}{\epsilon_1\sqrt{K^2+k_1^2} + \epsilon_2\sqrt{K^2+k_2^2}} \left(\int_0^\infty e^{-\sqrt{K^2+k_1^2}z}\bar{\rho}\left(\vec{K},z\right) dz\right)^2$$

For a spherical charged NP, the image potential energy can be calculated from the expression above by substituting in the right charge density.

$$\delta W = \frac{q^2}{32\pi\epsilon_0 \epsilon_1 a^2} \int_0^\infty \frac{KdK}{\sqrt{K^2+k_1^2}} \frac{\epsilon_1\sqrt{K^2+k_1^2} - \epsilon_2\sqrt{K^2+k_2^2}}{\epsilon_1\sqrt{K^2+k_1^2} + \epsilon_2\sqrt{K^2+k_2^2}} \left(\int_{z_0-a}^{z_0+a} e^{-\sqrt{K^2+k_1^2}z} J_0\left(K\sqrt{a^2-(z-z_0)^2}\right) dz\right)^2$$

For convenience, the image potential energy will be expressed in units of kT. Defining N as the number of elementary charges on the NP and $L_B = \frac{e^2}{4\pi\epsilon_0\epsilon_1 kT}$ as the Bjerrum length in dielectric 1, the expression above becomes:

$$\frac{\delta W}{kT} = \frac{N^2}{8a^2} L_B \int_0^\infty \frac{KdK}{\sqrt{K^2+k_1^2}} \frac{\epsilon_1\sqrt{K^2+k_1^2} - \epsilon_2\sqrt{K^2+k_2^2}}{\epsilon_1\sqrt{K^2+k_1^2} + \epsilon_2\sqrt{K^2+k_2^2}} \left(\int_{z_0-a}^{z_0+a} e^{-\sqrt{K^2+k_1^2}z} J_0\left(K\sqrt{a^2-(z-z_0)^2}\right) dz\right)^2$$

The inner integral can be rearranged to give a nicer expression.

$$I = \int_{z_0-a}^{z_0+a} e^{-\sqrt{K^2+k_1^2}z} J_0\left(K\sqrt{a^2-(z-z_0)^2}\right) dz \quad (A41)$$

A first step is to make the substitution $z = z_0 + at$

$$I = ae^{-\sqrt{K^2+k_1^2}z_0} \int_{-1}^{1} e^{-a\sqrt{K^2+k_1^2}t} J_0\left(Ka\sqrt{1-t^2}\right) dt$$

$$I = ae^{-\sqrt{K^2+k_1^2}z_0} \left[\int_{-1}^{0} e^{-a\sqrt{K^2+k_1^2}t} J_0\left(Ka\sqrt{1-t^2}\right) dt + \int_{0}^{1} e^{-a\sqrt{K^2+k_1^2}t} J_0\left(Ka\sqrt{1-t^2}\right) dt\right]$$

$$I = ae^{-\sqrt{K^2+k_1^2}z_0} \left[\int_{0}^{1} e^{a\sqrt{K^2+k_1^2}t} J_0\left(Ka\sqrt{1-t^2}\right) dt + \int_{0}^{1} e^{-a\sqrt{K^2+k_1^2}t} J_0\left(Ka\sqrt{1-t^2}\right) dt\right]$$

$$I = 2ae^{-\sqrt{K^2+k_1^2}z_0} \left[\int_{0}^{1} \cosh\left(a\sqrt{K^2+k_1^2}t\right) J_0\left(Ka\sqrt{1-t^2}\right) dt\right]$$

This form can be evaluated using an identity [c.f. Table of Integrals by Gradstein and Ryzhik.]

$$\int_0^a J_0\left(b\sqrt{a^2-x^2}\right)\cos(cx)\, dx = \frac{\sin\left(a\sqrt{b^2+c^2}\right)}{\sqrt{b^2+c^2}}, b > 0 \quad (A42)$$

Even if it is clear the identity is not of any use in this form, it can be adapted by making two useful substitutions: $x \to ax$ and $c \to ic$. After these substitutions the formula has to be adjusted for the case $b^2 < c^2$ because in our particular integral $K^2 < K^2 + k_1^2$ and leads to:

$$\int_0^1 J_0\left(ba\sqrt{1-x^2}\right)\cosh(cx) = \frac{\sinh\left(a\sqrt{c^2-b^2}\right)}{a\sqrt{c^2-b^2}} \quad (A43)$$

Applying this newly derived identity to our integral leads to the very nice formula:

$$I = 2ae^{-\sqrt{K^2+k_1^2}z_0}\frac{\sinh(k_1 a)}{k_1 a} \quad (A44)$$

Substituting it into our formula for the image potential energy gives the final result:

$$\boxed{\frac{\delta W}{kT} = N^2 \frac{L_B}{2}\left(\frac{\sinh(k_1 a)}{k_1 a}\right)^2 \int_0^\infty \frac{KdK}{\sqrt{K^2+k_1^2}} \frac{\epsilon_1\sqrt{K^2+k_1^2}-\epsilon_2\sqrt{K^2+k_2^2}}{\epsilon_1\sqrt{K^2+k_1^2}+\epsilon_2\sqrt{K^2+k_2^2}} e^{-2\sqrt{K^2+k_1^2}z_0}} \quad (A45)$$

Just for completeness, the self-energy of the spherical charge distribution will also be calculated. This is done by substituting $\tilde{\phi}_0$ in the expression for $W_0$.

$$W_0 = \frac{\pi^2}{\epsilon_0\epsilon_1}\int \frac{d\vec{K}}{\sqrt{K^2+k_1^2}}\int_0^\infty dz\tilde{\rho}\left(-\vec{K},z\right)\int_0^\infty e^{-\sqrt{K^2+k_1^2}|z-z'|}\tilde{\rho}\left(\vec{K},z'\right)dz' \quad (A46)$$

Substituting the charge density of a sphere in leads to:

$$W_0 = \frac{q^2}{32\pi\epsilon_0\epsilon_1 a^2}\int_0^\infty \frac{KdK}{\sqrt{K^2+k_1^2}}\int_{z_0-a}^{z_0+a} dzJ_0\left(K\sqrt{a^2-(z-z_0)^2}\right)\int_{z_0-a}^{z_0+a} dz' e^{-\sqrt{K^2+k_1^2}|z-z'|}J_0\left(K\sqrt{a^2-(z'-z_0)^2}\right)$$

### 1.3.3 Energy of a NP in the field created by a charged electrode:

Let us consider a charged electrode creates a potential $V_1\left(\vec{R},z\right)$ in a presence of any charge distribution $\rho\left(\vec{R},z\right)$. Then the energy imposed on the charge distribution by this potential is:

$$E = \int d\vec{R}\int_0^\infty dz V_1\left(\vec{R},z\right)\rho\left(\vec{R},z\right) \quad (A47)$$

For a spherical charge distribution of the system has two important features that help simplify the energy expression. First, the symmetry of the system makes both the potential and charge distribution depend only on R and z. Second, if we consider in reality NPs are not hollow, but made entirely of gold, the potential will stay constant across the NP. But first let us see how the energy formula changes when we substitute the charge density of a sphere.

$$E = 2\pi \int_0^\infty RdR \int_0^\infty dzV_1(R,z) \frac{q}{4\pi a^2} \delta\left(\sqrt{R^2 + (z-z_0)^2} - a\right)$$

We notice if R > a, the delta function becomes 0, so the integral over R can be evaluated only from 0 to a.

$$E = \frac{q}{2a^2} \int_0^a RdR \int_0^\infty dzV_1(R,z) \left[ \frac{\delta(z - z_0 + \sqrt{a^2 - R^2})}{\frac{\sqrt{a^2 - R^2}}{a}} + \frac{\delta(z - z_0 - \sqrt{a^2 - R^2})}{\frac{\sqrt{a^2 - R^2}}{a}} \right]$$

$$E = \frac{q}{2a} \int_0^a \frac{RdR}{\sqrt{a^2 - R^2}} \left[ V_1\left(R, z_0 - \sqrt{a^2 - R^2}\right) + V_1\left(R, z_0 + \sqrt{a^2 - R^2}\right) \right] (A48)$$

Now let us consider the potential generated by the electrode without the NP is given by $V(z)$. The next step is to relate $V_1(R,z)$ to $V(z)$. First we know the NP has radius a, so when R > a the two potentials coincide. When R < a the result is different. The potential remains the same as $V(z)$ until it enters the surface of the sphere, which happens through the point $z = z_0 - \sqrt{a^2 - R^2}$. Then it stays constant until it exits the surface, through the point $z_0 + \sqrt{a^2 - R^2}$. After this point the potential has the same values as $V(z)$, but shifted by the distance traveled through the sphere, which is $2\sqrt{a^2 - R^2}$. Knowing these details, we can construct an expression for $V_1(R,z)$ in terms of $V(z)$.

$$V_1(R,z) = \begin{cases} V(z), z \in \left[0, z_0 - \sqrt{a^2 - R^2}\right) \\ V\left(z_0 - \sqrt{a^2 - R^2}\right), z \in \left[z_0 - \sqrt{a^2 - R^2}, z_0 + \sqrt{a^2 - R^2}\right] (A49) \\ V\left(z - 2\sqrt{a^2 - R^2}\right), z \in \left(z_0 + \sqrt{a^2 - R^2}, \infty\right] \end{cases}$$

One physical requirement is to have a continuous potential, which is the case for $V_1$ as can be seen from the expression above.

Coming back to the energy, $V_1$ can be substituted to get:

$$E = \frac{q}{a} \int_0^a \frac{RdR}{\sqrt{a^2 - R^2}} V\left(z_0 - \sqrt{a^2 - R^2}\right) (A50)$$

A change of variable leads to:

$$E = \frac{q}{a} \int_0^a V(z_0 - t)dt$$

The next problem is to find out how the potential generated by the charged electrode varies with distance. Fortunately, it is described very well by the Gouy-Chapman theory, which assigns the following expression to the potential:

$$V(z) = \frac{2kT}{e} \ln \frac{\alpha + e^{-k_1 z}}{\alpha - e^{-k_1 z}} (A51)$$

Here $\alpha = \dfrac{e^{\frac{eV_0}{2kT}}+1}{e^{\frac{eV_0}{2kT}}-1}$ and $V_0$ is the potential of the electrode (the potential at $z = 0$).

$$E = \frac{q}{a}\frac{2kT}{e}\int_0^a \ln\frac{\alpha + e^{-k_1(z_0-t)}}{\alpha - e^{-k_1(z_0-t)}}dt$$

$$E = \frac{2NkT}{a}\int_0^a \ln\frac{1+\dfrac{e^{-k_1z_0}}{\alpha}e^{k_1t}}{1-\dfrac{e^{-k_1z_0}}{\alpha}e^{k_1t}}dt$$

Define a new parameter, $\beta = \dfrac{e^{-k_1z_0}}{\alpha}$

$$E = \frac{2NkT}{a}\int_0^a \ln\frac{1+\beta e^{k_1t}}{1-\beta e^{k_1t}}dt$$

Again, for convenience, the energy will be expressed in kT.

$$\boxed{\frac{E}{kT} = \frac{2N}{a}\int_0^a \ln\frac{1+\beta e^{k_1t}}{1-\beta e^{k_1t}}dt} \quad (A52)$$

### 1.3.4 Pair interaction potential energy:

In order to get the electrostatic potential energy between two NPs we have to first calculate the potential generated by one of them. Then the energy will be calculated by integrating this potential over the charge density of the other NP. Let us start by writing an expression for this energy, considering the center to center vector between the NPs is $\vec{R}_0$.

$$W = \int d^3\vec{r}\,\phi_1(\vec{r})\rho(\vec{r}-\vec{R}_0) \quad (A53)$$

$$W = \int d^3\vec{r}\,\phi_0(\vec{r})\rho(\vec{r}-\vec{R}_0) + \int d^3\vec{r}\,\delta\phi(\vec{r})\rho(\vec{r}-\vec{R}_0) \quad (A54)$$

Let us write

$$W = W_1 + W_2 \quad (A55)$$

$$W_1 = \int d^3\vec{r}\,\phi_0(\vec{r})\rho(\vec{r}-\vec{R}_0) \quad (A56)$$

$$W_2 = \int d^3\vec{r}\,\delta\phi(\vec{r})\rho(\vec{r}-\vec{R}_0) \quad (A57)$$

The reason for writing such general expressions for $W_1$ and $W_2$ is to be able to adapt them to any coordinate system. The notation $\phi_0$ was assigned to the value of the potential when no interface is present. Therefore, $\phi_0$ is spherically symmetric and the calculation of $W_1$ can be performed easily by using spherical coordinates. Regarding $W_2$, we know it is generated entirely by the presence of the interface, which generates cylindrical symmetry. So cylindrical coordinates can be used to evaluate $W_2$.

The first step is to calculate $\phi_0$. It can be written as an integral, using Green's function.

$$\phi_0(\vec{r}) = \frac{1}{4\pi\epsilon_0\epsilon_1} \int d^3\vec{r}' \frac{e^{-k_1|\vec{r}-\vec{r}'|}}{|\vec{r}-\vec{r}'|} \rho_1(\vec{r}') \quad (A58)$$

For a spherically symmetric system $\rho_1 = \rho_1(r)$, and the z axis can be aligned with $\vec{r}$ for easy evaluation.

$$\phi_0(r) = \frac{1}{2\epsilon_0\epsilon_1} \int_0^\infty r'^2 \rho_1(r') dr' \int_0^\pi \frac{e^{-k_1\sqrt{r^2+r'^2-2rr'\cos\theta}}}{\sqrt{r^2+r'^2-2rr'\cos\theta}} \sin\theta \, d\theta$$

$$\phi_0(r) = \frac{1}{2\epsilon_0\epsilon_1 r} \int_0^\infty r' \rho_1(r') dr' \int_0^\pi e^{-k_1\sqrt{r^2+r'^2-2rr'\cos\theta}} d\sqrt{r^2+r'^2-2rr'\cos\theta}$$

$$\phi_0(r) = \frac{1}{2\epsilon_0\epsilon_1 k_1 r} \int_0^\infty dr' r' \rho_1(r') \left(e^{-k_1|r-r'|} - e^{-k_1(r+r')}\right)$$

For a spherical NP with an overall charge of $q_1$ the charge density is

$$\rho_1(r') = \frac{q_1}{4\pi a^2} \delta(r'-a) \quad (A59)$$

Therefore, the potential becomes:

$$\phi_0(r) = \frac{q_1}{8\pi\epsilon_0\epsilon_1 k_1 ra} \left(e^{-k_1|r-a|} - e^{-k_1(r+a)}\right)$$

Considering NPs cannot overlap, we are only interested in the potential outside the sphere ($r \geq a$).

$$\boxed{\phi_0(r) = \frac{q_1}{4\pi\epsilon_0\epsilon_1} \frac{e^{-k_1 r}}{r} \frac{\sinh(k_1 a)}{k_1 a}} \quad (A60)$$

In order to make the calculations easier a quick change of variable is required for $W_1$, which is $\vec{r} \to \vec{r} + \vec{R}_0$.

$$W_1 = \int d^3\vec{r} \, \phi_0(\vec{r} + \vec{R}_0) \rho_2(\vec{r}) \quad (A61)$$

$$W_1 = \frac{q_1}{4\pi\epsilon_0\epsilon_1} \frac{\sinh(k_1 a)}{k_1 a} \int d^3\vec{r} \frac{e^{-k_1|\vec{r}+\vec{R}_0|}}{|\vec{r}+\vec{R}_0|} \rho_2(\vec{r})$$

Again, if $\rho_2 = \rho_2(r)$

$$W_1 = \frac{q_1}{2\epsilon_0\epsilon_1}\frac{sinh(k_1 a)}{k_1 a}\int_0^\infty r^2 \rho_2(r)dr \int_0^\pi \frac{e^{-k_1\sqrt{r^2+R_0^2+2rR_0\cos\theta}}}{\sqrt{r^2+R_0^2+2rR_0\cos\theta}}\sin\theta\, d\theta$$

$$W_1 = \frac{q_1}{2\epsilon_0\epsilon_1 R_0}\frac{sinh(k_1 a)}{k_1^2 a}\int_0^\infty r\rho_2(r)dr\left(e^{-k_1|r-R_0|} - e^{-k_1(r+R_0)}\right)$$

If the second NP has charge $q_2$ its charge density becomes

$$\rho_2(r) = \frac{q_2}{4\pi a^2}\delta(r-a) \quad (A62)$$

$$W_1 = \frac{q_1 q_2}{8\pi\epsilon_0\epsilon_1 R_0}\frac{sinh(k_1 a)}{k_1 a^3}\int_0^\infty r\left(e^{-k_1|r-R_0|} - e^{-k_1(r+R_0)}\right)\delta(r-a)$$

$$W_1 = \frac{q_1 q_2}{8\pi\epsilon_0\epsilon_1 R_0}\frac{sinh(k_1 a)}{k_1 a^2}\left(e^{-k_1|a-R_0|} - e^{-k_1(a+R_0)}\right)$$

We have to use again the fact that the NPs cannot overlap, so $R_0 \geq 2a$.

$$\boxed{W_1 = \frac{q_1 q_2}{4\pi\epsilon_0\epsilon_1}\frac{e^{-k_1 R_0}}{R_0}\left(\frac{sinh(k_1 a)}{k_1 a}\right)^2} \quad (A63)$$

Now to calculate $W_2$ we need to write it in cylindrical coordinates and in terms of the Hankel transforms of $\delta\phi$ and a shifted $\tilde{\rho}_2$, $\tilde{\rho}\left(\vec{K},z,\vec{R}_0\right)$.

$$W_2 = 4\pi^2 \int d\vec{K}\int_0^\infty dz\, \widetilde{\delta\phi}\left(\vec{K},z\right)\tilde{\rho}\left(-\vec{K},z,\vec{R}_0\right) \quad (A64)$$

Calculating $\widetilde{\delta\phi}$:

$$\widetilde{\delta\phi}\left(\vec{K},z\right) = \frac{1}{2\epsilon_0\epsilon_1\sqrt{K^2+k_1^2}}\frac{\epsilon_1\sqrt{K^2+k_1^2}-\epsilon_2\sqrt{K^2+k_2^2}}{\epsilon_1\sqrt{K^2+k_1^2}+\epsilon_2\sqrt{K^2+k_2^2}}\int_0^\infty e^{-\sqrt{K^2+k_1^2}(z+z')}\tilde{\rho}_1\left(\vec{K},z'\right)dz'$$

$$\widetilde{\delta\phi}\left(\vec{K},z\right)$$
$$= \frac{q_1}{16\pi^2 a\epsilon_0\epsilon_1\sqrt{K^2+k_1^2}}\frac{\epsilon_1\sqrt{K^2+k_1^2}-\epsilon_2\sqrt{K^2+k_2^2}}{\epsilon_1\sqrt{K^2+k_1^2}+\epsilon_2\sqrt{K^2+k_2^2}}e^{-\sqrt{K^2+k_1^2}z}\int_{z_0-a}^{z_0+a}e^{-\sqrt{K^2+k_1^2}z'}J_0\left(K\sqrt{a^2-(z'-z_0)^2}\right)dz'$$

$$\widetilde{\delta\phi}\left(\vec{K},z\right) = \frac{q_1}{8\pi^2\epsilon_0\epsilon_1}\frac{e^{-\sqrt{K^2+k_1^2}(z+z_0)}}{\sqrt{K^2+k_1^2}}\frac{\epsilon_1\sqrt{K^2+k_1^2}-\epsilon_2\sqrt{K^2+k_2^2}}{\epsilon_1\sqrt{K^2+k_1^2}+\epsilon_2\sqrt{K^2+k_2^2}}\frac{sinh(k_1 a)}{k_1 a} \quad (A65)$$

$$\rho\left(\vec{R},z,\vec{R}_0\right) = \rho_2\left(\vec{R}-\vec{R}_0,z\right) \quad (A66)$$

$$\tilde{\rho}\left(\vec{K},z,\vec{R}_0\right) = \frac{1}{(2\pi)^2}\int d\vec{R}\, e^{-i\vec{K}\cdot\vec{R}}\rho_2\left(\vec{R}-\vec{R}_0,z\right)$$

Change of variable: $\vec{R} \to \vec{R} + \vec{R}_0$

$$\tilde{\rho}\left(\vec{K}, z, \vec{R}_0\right) = \frac{1}{(2\pi)^2} e^{-i\vec{K}\cdot\vec{R}_0} \int d\vec{R} e^{-i\vec{K}\cdot\vec{R}} \rho_2\left(\vec{R}, z\right)$$

$$\tilde{\rho}\left(\vec{K}, z, \vec{R}_0\right) = e^{-i\vec{K}\cdot\vec{R}_0} \tilde{\rho}_2\left(\vec{K}, z\right)$$

$$\tilde{\rho}\left(\vec{K}, z, \vec{R}_0\right) = e^{-i\vec{K}\cdot\vec{R}_0} \frac{q_2}{8\pi^2 a} J_0\left(K\sqrt{a^2 - (z-z_0)^2}\right)$$

$W_2$

$$= \frac{q_1 q_2}{16\pi^2 \epsilon_0 \epsilon_1 a} \frac{\sinh(k_1 a)}{k_1 a} \int d\vec{K} \frac{e^{-\sqrt{K^2+k_1^2}z_0}}{\sqrt{K^2+k_1^2}} \frac{\epsilon_1\sqrt{K^2+k_1^2} - \epsilon_2\sqrt{K^2+k_2^2}}{\epsilon_1\sqrt{K^2+k_1^2} + \epsilon_2\sqrt{K^2+k_2^2}} e^{-i\vec{K}\cdot\vec{R}_0} \int_{z_0-a}^{z_0+a} dz\, e^{-\sqrt{K^2+k_1^2}z} J_0\left(K\sqrt{a^2 - (z-z_0)^2}\right)$$

$$W_2 = \frac{q_1 q_2}{8\pi^2 \epsilon_0 \epsilon_1} \left(\frac{\sinh(k_1 a)}{k_1 a}\right)^2 \int d\vec{K} \frac{e^{-2\sqrt{K^2+k_1^2}z_0}}{\sqrt{K^2+k_1^2}} \frac{\epsilon_1\sqrt{K^2+k_1^2} - \epsilon_2\sqrt{K^2+k_2^2}}{\epsilon_1\sqrt{K^2+k_1^2} + \epsilon_2\sqrt{K^2+k_2^2}} e^{-i\vec{K}\cdot\vec{R}_0}$$

$$W_2 = \frac{q_1 q_2}{8\pi^2 \epsilon_0 \epsilon_1} \left(\frac{\sinh(k_1 a)}{k_1 a}\right)^2 \int_0^\infty \frac{K dK}{\sqrt{K^2+k_1^2}} \frac{\epsilon_1\sqrt{K^2+k_1^2} - \epsilon_2\sqrt{K^2+k_2^2}}{\epsilon_1\sqrt{K^2+k_1^2} + \epsilon_2\sqrt{K^2+k_2^2}} e^{-2\sqrt{K^2+k_1^2}z_0} \int_{-\pi}^{\pi} d\theta\, e^{-KR_0 \cos\theta}$$

$$\boxed{W_2 = \frac{q_1 q_2}{4\pi \epsilon_0 \epsilon_1} \left(\frac{\sinh(k_1 a)}{k_1 a}\right)^2 \int_0^\infty \frac{K dK}{\sqrt{K^2+k_1^2}} \frac{\epsilon_1\sqrt{K^2+k_1^2} - \epsilon_2\sqrt{K^2+k_2^2}}{\epsilon_1\sqrt{K^2+k_1^2} + \epsilon_2\sqrt{K^2+k_2^2}} e^{-2\sqrt{K^2+k_1^2}z_0} J_0(KR_0)} \quad (A67)$$

Now the total pair energy can be written as:

$$\boxed{W = \frac{q_1 q_2}{4\pi \epsilon_0 \epsilon_1} \left(\frac{\sinh(k_1 a)}{k_1 a}\right)^2 \left(\frac{e^{-k_1 R_0}}{R_0} + \int_0^\infty \frac{K dK}{\sqrt{K^2+k_1^2}} \frac{\epsilon_1\sqrt{K^2+k_1^2} - \epsilon_2\sqrt{K^2+k_2^2}}{\epsilon_1\sqrt{K^2+k_1^2} + \epsilon_2\sqrt{K^2+k_2^2}} e^{-2\sqrt{K^2+k_1^2}z_0} J_0(KR_0)\right)} \quad (A68)$$

Again, for convenience this energy will be expressed in units of kT:

$$\boxed{\frac{W}{kT} = N_1 N_2 L_B \left(\frac{\sinh(k_1 a)}{k_1 a}\right)^2 \left(\frac{e^{-k_1 R_0}}{R_0} + \int_0^\infty \frac{K dK}{\sqrt{K^2+k_1^2}} \frac{\epsilon_1\sqrt{K^2+k_1^2} - \epsilon_2\sqrt{K^2+k_2^2}}{\epsilon_1\sqrt{K^2+k_1^2} + \epsilon_2\sqrt{K^2+k_2^2}} e^{-2\sqrt{K^2+k_1^2}z_0} J_0(KR_0)\right)} \quad (A6$$

## 2. Drude-Lorentz model for Au and TiN

For completeness, theoretical curves for dielectric constants are also compared to literature data for both Au[33] and TiN.[37]

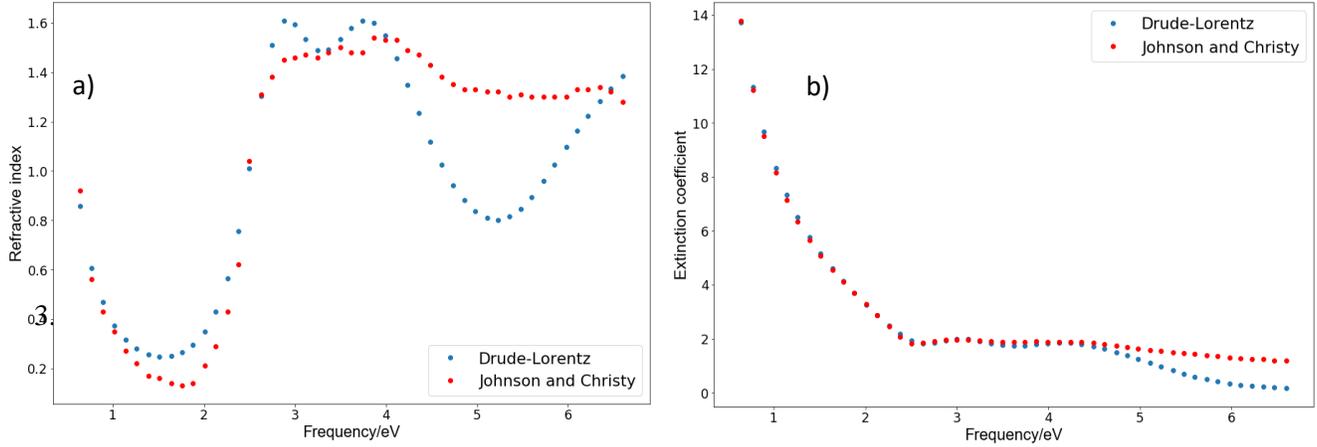

Fig. 1 a) Refractive index of Au, comparison between Johnson and Christy's dataset and Drude-Lorentz model. b) Extinction coefficient of Au, comparison between Johnson and Christy's dataset and Drude-Lorentz model.

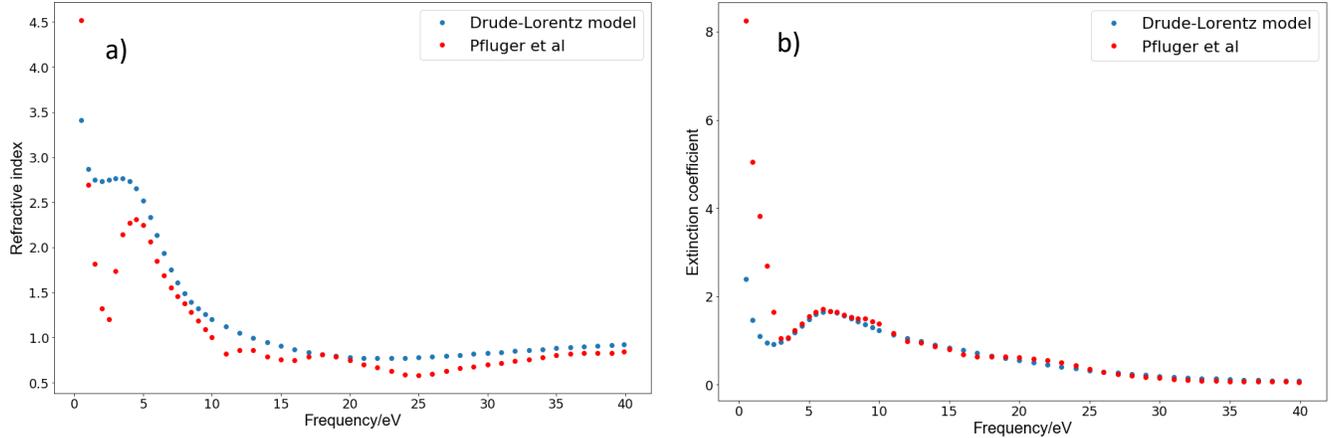

Fig. 2 a) Refractive index of TiN, comparison between Pfluger's dataset and Drude-Lorentz model. b) Extinction coefficient of TiN, comparison between Pfluger's dataset and Drude-Lorentz model.